\documentclass[%
 aip,
 amsmath,amssymb,
 reprint,%
]{revtex4-1}

\usepackage{float}
\usepackage{graphicx}% Include figure files
\usepackage{dcolumn}% Align table columns on decimal point
\usepackage{bm}% bold math

\usepackage{xcolor}
\usepackage[normalem]{ulem}
\usepackage{siunitx}
\usepackage{upgreek}
\usepackage{appendix}

\usepackage[utf8]{inputenc}
\usepackage[T1]{fontenc}
\usepackage{mathptmx}

\newcommand{\trm}[1]{\textrm{#1}}
\newcommand{\prens}[1]{\left(#1\right)}

	\newcommand{\perm}{\varepsilon}
	\newcommand{\U}{\mathbf{u}}
	\newcommand{\E}{\mathbf{E}}
	\newcommand{\gsquig}{g}
	
	\newcommand{\proj}{\mathbf{\Pi}}
\newcommand{\tbmu}{t_{\trm{b}\upmu}}

\definecolor{cjsblue}{HTML}{4b50a8}

\definecolor{slate}{HTML}{708090}

\renewcommand{\proj}{\Pi}

\begin{document}
\title[Acousto-optic modulation in lithium niobate on sapphire]{Acousto-optic modulation in lithium niobate on sapphire}

\author{Christopher J. Sarabalis}
\email{sicamor@stanford.edu}
\author{Timothy P. McKenna}
\author{Rishi N. Patel}
\author{Rapha\"{e}l Van Laer}
\author{Amir H. Safavi-Naeini}
\email{safavi@stanford.edu}

\affiliation{%
 Department of Applied Physics and Ginzton Laboratory, Stanford University\\
 348 Via Pueblo Mall, Stanford, California 94305, USA
}%
%\altaffiliation[Also at ]{}%Lines break automatically or can be forced with \\

\date{\today}% It is always \today, today,

\begin{abstract}
	We demonstrate acousto-optic phase modulators in X-cut lithium niobate films on sapphire, detailing the dependence of the piezoelectric and optomechanical coupling coefficients on the crystal orientation. This new platform supports highly confined, strongly piezoelectric mechanical waves without suspensions, making it a promising candidate for broadband and efficient integrated acousto-optic devices, circuits, and systems.

\end{abstract}

\maketitle

The demonstration of low-loss nanophotonic waveguides~\cite{Wang2014} in high-quality, single-crystal films of lithium niobate (LN)~\cite{Levy1998} has led to a surge in the development of electro-optic and nonlinear optical devices. 
Because of their small mode area, these waveguides exhibit large nonlinear interactions and require less energy to parametrically drive, underlying recent progress in LN frequency combs~\cite{Yu2019}, second-harmonic generation~\cite{Rao2016, Chang2016}, and high-speed electro-optic modulators~\cite{Rao2016b, Wang2018}.

In addition to the large \(\chi^{\prens{2}}\) nonlinearity and electro-optic effect that make LN an attractive material for optics, LN is a low-loss mechanical material with strong piezoelectric coupling, properties that are necessary for making broadband and efficient acousto-optic modulators (AOMs).  In parallel to the development of nanophotonics in thin film LN, strongly piezoelectric, low-loss resonators~\cite{Olsson2014,Pop2017} and delay-lines~\cite{Manzaneque2019} have been demonstrated and wavelength-scale waveguides efficiently transduced~\cite{Dahmani2019} in suspended films. Using these piezoelectric devices, suspended LN AOMs have realized low-power microwave-to-optical conversion in pursuit of quantum optical interconnects~\cite{Shao2019, Jiang2019}. 

While suspended devices are at the forefront of low-power acousto-optics, suspensions add fabrication constraints that inhibit the development of complex photonic and phononic circuits and systems. Another approach to co-localize optical and mechanical waves is to employ the Rayleigh-like surface acoustic wave (SAW) which is confined to the surface in any material platform. These SAWs can modulate a variety of optical structures including resonators~\cite{Tadesse2014}; waveguides as recently used to demonstrate non-magnetic isolation~\cite{Kittlaus2020}; Mach-Zehnders for intensity modulators~\cite{VanDerPoel2007,Slot2019, Cai2019} and acousto-optic gryoscopes~\cite{Mahmoud2018, Mahmoud2018b}; and arrayed waveguide gratings~\cite{Poveda2015,Poveda2016} with Refs.~\cite{Mahmoud2018,Mahmoud2018b,Cai2019} using LN-on-insulator. But LN-on-insulator has a fundamental drawback. Analogous to the advantage of high confinement in electro-optics and nonlinear optics, high mechanical confinement is necessary for broadband electromechanical transduction as well as for efficient optomechanical modulation. In LN-on-insulator, however, as the mechanical wavelength approaches the thickness of the LN film, the mechanical wave leaks out of the LN and into the silica which supports slower mechanical waves. This can be addressed by replacing the insulator substrate with a material with higher sound velocities enabling high-confinement mechanical waves in the thin-film device layer.

Here we explore integrated acousto-optic modulation in X-cut, thin-film LN bonded to sapphire with a focus on verifying the piezoelectric and acousto-optic properties of our film. This new platform enables high-confinement optical waveguides from the near-infrared to the visible and, in addition to Rayleigh-like SAWs, supports guided horizontal shear (SH) waves that exhibit large electromechanical coupling coefficients \(k_\trm{eff}^2\) at GHz-frequencies. Compared to related efforts in other platforms making use of aluminum nitride, silicon, or gallium arsenide, this platform enables strong confinement of the mechanical waves without suspensions, setting the stage for complex phononic circuitry and systems. We demonstrate a surface wave acousto-optic phase modulator (discussed in Section~\ref{sec:deviceAndNumerics}) utilizing the Rayleigh and SH modes near $700$ and \SI{800}{\mega\hertz}, respectively, and characterize them in the telecom C-band. By comparing simulations and measurements of the piezoelectric coupling coefficient \(k_\trm{eff}\) (Section~\ref{sec:mechanics}) and the optomechanical coupling coefficient \(\gsquig\) (Section~\ref{sec:AO}), we show the degree to which these bonded films retain their piezoelectric and acousto-optic properties. The acousto-optic portion of the study is similar to recent work by Khan~\emph{et al.} which extracted the dominant elasto-optic coefficients for waveguides patterned in sputtered arsenic trisulfide films~\cite{Khan2019}. Furthermore, we show that at GHz frequencies as the wavelength approaches the LN film thickness, the piezoelectric coupling of the SH wave quickly increases with \(k_\trm{eff}^2\) exceeding $10\%$ just above \SI{2}{\giga\hertz}.

\section{Modeling SAW phase modulators}
\label{sec:deviceAndNumerics}

\begin{figure}[h]
    \centering
    \includegraphics{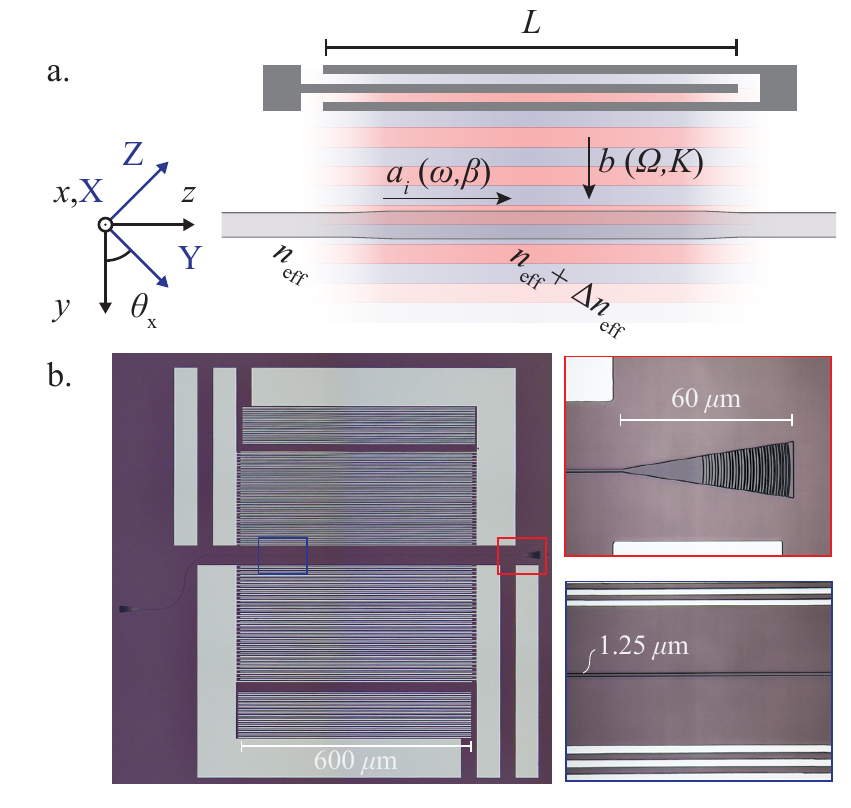}
    \caption{\textbf{SAW Phase Modulator.} \textbf{a.} Aluminum IDTs generate surface waves in the LN film on sapphire which travel normal to an optical ridge waveguide. These surfaces waves modulate the effective index of the optical modes in the waveguide. \textbf{b.} Pictures of the fabricated device showing the IDTs, ridge waveguide, and focusing grating couplers.}
    \label{fig:device}
\end{figure}

The surface wave phase modulator shown in Figure~\ref{fig:device}a is a simple acousto-optic device with two parts : a piezoelectric transducer to generate mechanical waves and an optical ridge waveguide modulated by these waves. Surface waves are generated by an interdigitated transducer (IDT) with phase fronts parallel to an optical ridge waveguide. These surface waves modulate the effective index of refraction \(n_\trm{eff}\) of the waveguide and therefore the phase of light transmitted through the device. 

The piezoelectric transducer is characterized by two numbers, the effective piezoelectric coupling coefficient \(k_\trm{eff}\) and the transmission coefficient \(\tbmu\) from microwaves incident on the IDT to phonons in a specific mechanical mode and direction~\cite{Hashimoto2000,Hashimoto2009,Dahmani2019}.  For our purposes, the most important characteristic of the ridge waveguide is the optomechanical coupling coefficient \(\gsquig\) which has units of \(\prens{\trm{W}\cdot\trm{m}}^{-1/2}\) and is defined in Appendix~\ref{app:OM}. Of these figures, \(k_\trm{eff}\) and \(\gsquig\) are proportional to the piezoelectric and photoelastic tensors~\cite{Andrushchak2009}, respectively, and so can be used to characterize the quality of the bonded film and platform. For this reason, they are the focus of our study. But in order to extract \(\gsquig\) from optical measurements of the modulation index \(h_\trm{ao}\), we also need to determine \(\tbmu\).

\begin{figure*}[t]
    \centering
    \includegraphics{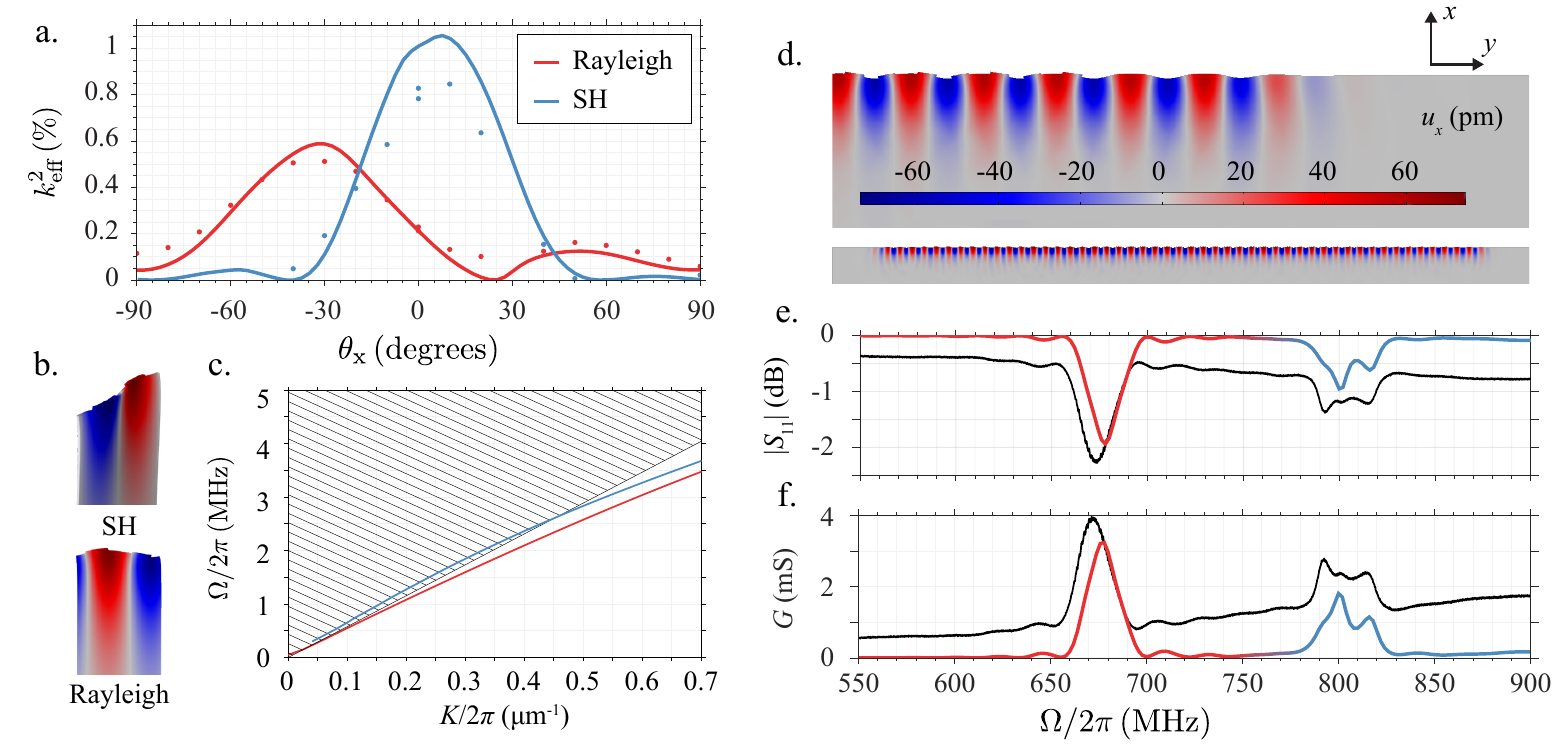}
		\caption{\textbf{Piezoelectric response.} \textbf{a}. We compute the piezoelectric coupling coefficient \(k_\trm{eff}^2\) from the X-point solutions for a unit cell of the IDT (\textbf{b.}). The Rayleigh wave (red) and SH wave (blue) exhibit \(k_\trm{eff}^2\) on the order of \(1\%\). Measurements of \(k_\trm{eff}^2\) (points) are overlaid on the simulated curves. \textbf{c.} Bands for the slab (no electrodes) with \(\theta_\trm{x} = 0^\circ\) show the Rayleigh mode (red) is guided, \emph{i.e.} below the sound cone in sapphire (hatched), for all frequencies. At \SI{2.7}{\giga\hertz}, the SH mode (blue) emerges from the sound cone, but even below this frequency, the radiation-limited propagation length is long enough to use the mode for modulation. \textbf{d}. The conductance is computed for a thin cross-section of the transducer. The \(S_{11}\), \textbf{e}., and conductance \(G\), \textbf{f}., are plotted for measurements (black) and simulations (red-blue gradient colored by mode).}
    \label{fig:mechanics}
\end{figure*}

First we consider numerical analyses of the transducer in Figure~\ref{fig:device}b, focusing on \(k_\trm{eff}\) before considering \(\tbmu\). A large \(k_\trm{eff}\) is essential for making small, broadband transducers~\cite{Dahmani2019}. The coupling can be estimated efficiently from quantities computed on a unit cell of an IDT, specifically, the series and parallel resonance frequencies, \(\Omega_\trm{s}\) and \(\Omega_\trm{p}\). We simulate a thin, three-dimensional cross-section of a finger pair with Floquet boundary conditions along the direction of propagation \(\hat{y}\) (see Figure~\ref{fig:mechanics}b). We assume continuity along \(\hat{z}\).
In a lossless simulation, the series and parallel resonances correspond to short and open boundary conditions across the electrodes. To first order in \((\Omega_\trm{p} - \Omega_\trm{s})/\Omega_\trm{p}\)~\cite{Dahmani2019},
\begin{equation}
    k_\trm{eff}^2 = \frac{\pi^2}{8}\prens{\frac{\Omega_\trm{p}^2}{\Omega_\trm{s}^2} - 1}.
\end{equation}
A \SI{225}{nm}-thick LN slab on sapphire supports a Rayleigh-like mode and a leaky horizontal shear (SH) mode with a wavelength of \SI{8}{\micro\meter} and frequencies near \SI{750}{MHz}. The coupling \(k_\trm{eff}^2\) for these modes depends on the orientation of the electrodes with respect to the extraordinary axis as plotted in Figure~\ref{fig:mechanics}a. 

The level of confinement of the acoustic wave depends strongly on its wavelength $\Lambda$ and therefore its frequency. At \(\Lambda = \SI{8}{\micro\meter}\) for the SH mode, three quarters of the mechanical energy is in the sapphire substrate and so \(k_\trm{eff}^2\) only reaches 1.2\%. At shorter wavelengths, more energy is confined to the LN film and the coupling increases. We show in Section~\ref{sec:mechanics} that \(k_\trm{eff}^2\) of the SH waves exceeds 10\% for \SI{2}{\micro\meter}-pitch IDTs where 32\% of the energy is in the LN (with 39\% in the sapphire and 29\% in the electrodes).  By comparison, the \(k_\trm{eff}^2\)  in suspended LN films reaches 30\%~\cite{Pop2017}. Furthermore at \SI{8}{\micro\meter}, the SH band (blue in Figure~\ref{fig:mechanics}c) is phase-matched to waves in the sapphire (hatched) and so the wave leaks into the substrate at a rate of $10~\trm{dB}/\trm{mm}$. Above \SI{2.7}{\giga\hertz}, the SH wave is no longer leaky. Smaller $\Lambda$ will be pursued in future work to achieve higher confinement and lower acoustic radiation loss.

In order to determine the coupling coefficient \(\gsquig\) from measurements of the modulation index (Section~\ref{sec:AO}), we need the efficiency of the IDT \(\tbmu\).   This coefficient can be expressed as the product of two factors~\cite{Dahmani2019} 
\begin{equation}
	\left|\tbmu\right|^2 = \prens{1 - |S_{11}|^2}\frac{|a_i|^2}{2 G V^2}.
	\label{eq:tbmu}
\end{equation}
The first factor comes from impedance mismatch and captures the fraction of incident microwave power that gets reflected. It can be determined experimentally from measurements of \(S_{11}\), the one-port S-parameter of the IDT. It can also be calculated from the admittance of solutions of the inhomogeneous piezoelectric equations of the full IDT and reflector bounded by perfectly matched absorbing layers (Figure~\ref{fig:mechanics}d). The second factor in Equation~\ref{eq:tbmu} is the fraction of power radiated into the \(i\)th mode which is computed by decomposing the radiation into a basis of waves in the slab~\cite{Auld1990v2}. The amplitude \(a_i\) is normalized such that \(\left|a_i\right|^2\) is the power in mode \(i\). The product \(2 G V^2\) in the denominator is the total power dissipated when a voltage \(V\) is applied accross the IDT. Details on these methods are presented by Dahmani~\emph{et al.}~\cite{Dahmani2019}.

The surface waves generated by the IDT strain and deform the optical waveguides. This changes the effective index \(n_\trm{eff}\) of the fundamental TE-like optical mode as captured by the optomechanical coupling coefficient \(\gsquig\). The coupling coefficient is computed from the optical and piezoelectric eigenmodes of an extruded cross-section of the waveguide solved for by FEM in COMSOL~\cite{COMSOL} to capture the full vectorial nature of the fields. The optical mode propagates into the plane in Figure~\ref{fig:AO}a and the piezoelectric mode across the plane in Figure~\ref{fig:AO}b. The electric field of the TE-like optical mode is antisymmetric with respect to the \(xz\) symmetry plane. At each \(\theta_\trm{x}\), these solutions are used to evaluate \(\gsquig\) by the perturbative overlap integral (detailed in Appendix~\ref{app:OM})
\begin{equation}
	\gsquig = -\frac{\omega}{\mathcal{P}_1} \int\trm{d}A \E^*_1\delta_u\perm\cdot \U_0 \E_0.
	\label{eq:gsquig_main_body}
\end{equation}
$\E_j$ and $\U_k$ are the electromagnetic and displacement field distributions for the guided mode solutions, \(\omega\) is the optical frequency, \(\mathcal{P}_j\) is the time-averaged optical power, and $\delta_u\perm$ is the modification of the structure's dielectric constant distribution due to motion, which includes both the shifts in the boundaries and the contribution of the photoelastic tensor.

Despite the nontrivial dependence of \(\gsquig\) on waveguide orientation as plotted  in Figure~\ref{fig:AO}, a simple picture describes the interaction at the peaks. Consider the Rayleigh and fundamental TE modes. The \(S_{yy}\) component of the strain --- the dominant component of the Rayleigh waves at the surface --- modulates the \(\perm_{yy}\) component of the permittivity via the \(p_{yyyy}\) component of the photoelastic tensor. Modulating \(\perm_{yy}\) modulates the TE mode which has a \(y\)-oriented electric field. In Figure~\ref{fig:AO}c, the coupling coefficient peaks at \(\theta_\trm{x}= -45^\circ\) where \(p_{yyyy}\) reaches a maximum of \(0.30\) for X-cut LN.  Similarly, interactions with the SH mode are dominated by \(p_{yyyz}\) which has local extrema of \(0.127\) and \(-0.135\)  at \(\theta_\trm{x} = -18^\circ\) and \(32^\circ\), respectively.

\section{Fabrication and characterization}

We start our process with \(5\times\SI{10}{\milli\meter}\) chips of \SI{525}{\nano\meter}-thick LN-on-sapphire. The LN is X-cut and the c-axis of the sapphire is normal to the wafer.  The a-axis of the sapphire and the Z-axis of the LN are in-plane and parallel. Ridge waveguides and grating couplers are patterned into a hydrogen silsesquioxane (HSQ) mask and transferred to the sample by a $\SI{300}{\nano\meter}$ argon ion etch leaving a $\SI{225}{\nano\meter}$ thick LN slab on the sapphire substrate. The remaining mask is stripped with hydrofluoric acid before the chip is cleaned with piranha.  The $\SI{200}{\nano\meter}$ thick aluminum electrodes are patterned by lift-off on the \SI{225}{\nano\meter} LN slab. The IDT is \(\SI{600}{\micro\meter}\) wide and has \(30\) aluminum finger-pairs. The ridge waveguide is \SI{1.25}{\micro\meter} wide and supports a TE-like and a TM-like optical mode.

Below we describe how \(k_\trm{eff}\) and \(\gsquig\) are extracted from measurements of the IDTs' linear response and the modulators' modulation index.

\subsection{Piezoelectric coupling coefficient in LN-on-sapphire}
\label{sec:mechanics}

In order to characterize the piezoelectric quality of the film, we measure the coupling coefficient \(k_\trm{eff}^2\) and compare it to simulations. The coupling coefficient is extracted from measurements of the one-port microwave response \(S_{11}\) of the IDT for a range of crystal orientations \(\theta_\trm{x}\) varying from \(-90^\circ\) to \(90^\circ\). 

We measure the S-parameter of each device on a calibrated probe station (GGB 40-A nickel probes) with an R\&S ZNB20 vector network analyzer (Figure~\ref{fig:mechanics}e) to determine the admittance \(Y\prens{\omega}\). The coupling can be computed directly from the measured conductance \(G(\omega)~\equiv~\trm{Re}Y(\omega)\)~\cite{Dahmani2019}
\begin{equation} 
	k_\trm{eff}^2 = \frac{\pi\int\trm{d}\omega G\prens{\omega}}{4 C_0 \Omega_0^2}
	\label{eq:ksquaredByNetConductance}
\end{equation}
where \(C_0\) is the static capacitance computed by fit to the susceptance \(-\trm{Im}Y(\omega)\) near DC, \(\Omega_0\) is the center frequency of the response, and the integral is evaluated about \(\Omega_0\). In addition to the mechanical signature (red-blue curve in Figure~\ref{fig:mechanics}f), $G(\omega)$ is offset by a slowly increasing background which comes from ohmic loss and inductance of the IDT. This effect is not captured in the cross-section modeled in Section~\ref{sec:deviceAndNumerics} which assumes the fields are uniform along the fingers. We fit the background to a parabola and  remove it from \(G\) before computing \(k_\trm{eff}^2\) by Equation~\ref{eq:ksquaredByNetConductance}. This gives us the points in Figure~\ref{fig:mechanics}a.

We find excellent agreement between the shape of the simulated and measured \(k_\trm{eff}^2(\theta_\trm{x})\) as shown in Figure~\ref{fig:mechanics}a. The magnitude of the Rayleigh response matches with simulation, but the SH response falls 20\% below the simulated response at its peak near \(0^\circ\). A reduction of around $10\%$ in the piezoelectric tensor component \(d_\trm{YZY}\) from $69$ to \SI{62}{\pico\coulomb\per\newton} would lead to this reduction in $k_\trm{eff}^2$.

\begin{figure}[h]
    \centering
    \includegraphics{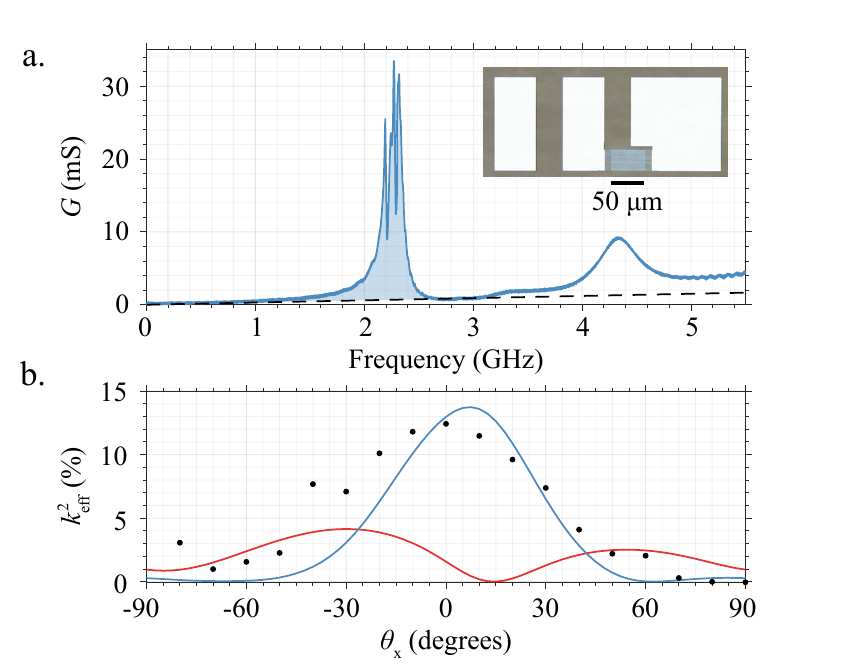}
    \caption{\textbf{High \(k_\trm{eff}^2\) mechanics.} At higher frequencies, the coupling coefficient of the SH mode exceeds $10\%$. \textbf{a.} The conductance of \SI{2}{\micro\meter}-pitch IDTs can be integrated near the Rayleigh and SH response just above \SI{2}{\giga\hertz} to calculate the sum of \(k_\trm{eff}^2\) for the two modes. A linear background (dashed) is removed from $G$. An example device is inset. \textbf{b.} Measurements of the summed \(k_\trm{eff}^2\) (black dots) are plotted against simulations of the coupling of Rayleigh (red) and SH (blue) mode.}
    \label{fig:GHzSH}
\end{figure}

An important advantage of the LN-on-sapphire platform is that horizontal shear (SH) waves are strongly piezoelectric at high frequency. To demonstrate this, we repeat the above procedure for \SI{2}{\micro\meter}-pitch IDTs on \SI{200}{\nano\meter}-thick LN-on-sapphire. A typical conductance curve is shown in Figure~\ref{fig:GHzSH}a for \(\theta_\trm{x} = 0^\circ\). For these $15$-finger-pair, \SI{50}{\micro\meter}-wide transducers, the proximity of the Rayleigh and SH waves makes it difficult to independently filter their contributions to \(G\). Instead we integrate the conductance for the blue shaded region (between $1$ and \SI{2.65}{\giga\hertz}) and report the sum of \(k_\trm{eff}^2\) for both modes (Figure~\ref{fig:GHzSH}b). Simulations show that the Rayleigh mode remains weakly coupled, and so the increase in \(k_\trm{eff}^2\) is primarily due to the SH mode. The results of measurements of IDTs at a variety of angles are plotted in Figure~\ref{fig:GHzSH}, showing that \(k_\trm{eff}^2\) for the SH mode exceeds $10\%$.

\subsection{Optomechanics in LN-on-sapphire}
\label{sec:AO}

We investigate the optomechanical properties of the bonded LN film --- in particular, the photoelastic coefficients --- by measuring the optomechanical coupling coefficient \(\gsquig\) of a nanophotonic ridge waveguide and comparing it to simulations. As with \(k_\trm{eff}^2\), \(\gsquig\) varies with crystal orientation because LN is strongly anisotropic.

\begin{figure}
    \centering
    \includegraphics{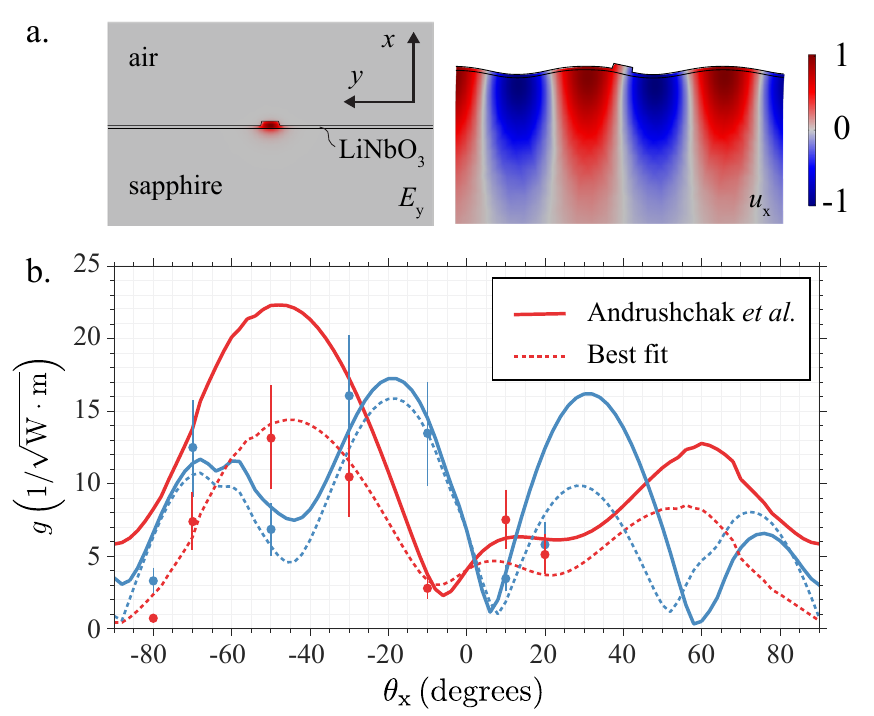}
    \caption{\textbf{Optomechanical response.} \textbf{a}. The effective index of the TE optical mode (left) is modulated by the Rayleigh wave (right). \textbf{b}. Angular dependence of the simulated (solid, photoelastic tensor from Ref.~\cite{Andrushchak2009}) and measured (dots) optomechanical coupling coefficient. Simulations of \(g\) with best fit photoelastic coefficients are plotted (dashed). Red and blue are Rayleigh and SH modes as before.}
    \label{fig:AO}
\end{figure}

\begin{figure}
    \centering
    \includegraphics[width=\linewidth]{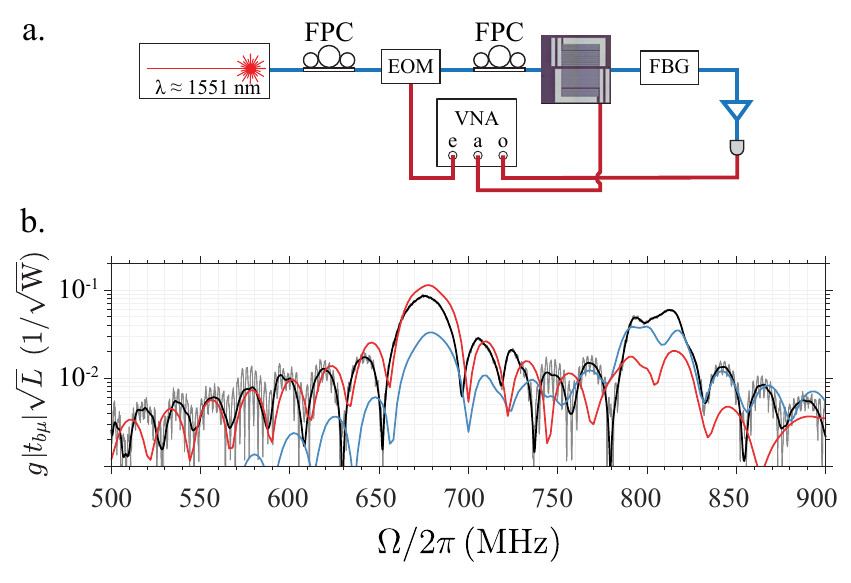}
    \caption{\textbf{Acousto-optic measurements.} \textbf{a}. Apparatus used to measure the modulation index. \textbf{b}. Measurements of the modulation efficiency \(h_\trm{ao}/2\sqrt{P}\) in gray are overlaid with predictions \(\gsquig\tbmu\sqrt{L}\) from FEM simulations of the Rayleigh (red) and SH (blue) modes for \(\theta_\trm{x} = -30^\circ\). The black curve is filtered in the time-domain to eliminate contributions from reflections.}
    \label{fig:Seo}
\end{figure}

We determine $\gsquig$ by sending light through the device while driving the IDT with a microwave signal, and measuring the modulation index $h_\trm{oa}$ of the transmitted light. The IDT transmission \(\tbmu\) is used in this calibration to determine the mechanical power incident on the waveguide due to the microwave driving. We measure the modulation index with the apparatus diagrammed in Figure~\ref{fig:Seo}a which is used to detect the resulting phase modulation. We tune a C-band laser (Santec TSL-550) to the edge of a Teraxion fiber Bragg notch filter near $1551$~nm such that phase modulation by the device is converted to intensity modulation. This intensity modulated light is then amplified (FiberPrime EDFA-C-26G-S) and detected on a photodiode (Optilab PD-40-M). We drive the modulator and read out the photocurrent fluctuations on a vector network analyzer (VNA, R\&S ZNB20). The modulation index \(h_\trm{oa}\) is calibrated by comparing the acousto-optic signal to phase modulation from an electro-optic modulator (EOM) cascaded with the device
\begin{equation}
    h_\trm{oa} = h_\trm{oe}\left|\frac{S_\trm{oa}}{S_\trm{oe}}\right|.
\end{equation}
Here \(S_\trm{oa}\) and \(S_\trm{oe}\) are S-parameters measured on the VNA for an acousto-optic device and the EOM, respectively, at the peak acousto-optic response. The EOM is calibrated independently using a tunable Fabry-P\'erot filter (Micron Optics FFP-TF) to filter and measure the power of the pump and sidebands for a given RF drive power \(P\). 

The modulation index \(h_\trm{oa}\) is not a direct measurement of the coupling coefficient. Instead the measured quantity plotted in Figure~\ref{fig:Seo}b 
\begin{equation}
	\frac{h_\trm{oa}}{2\sqrt{P}} = \gsquig\left|\tbmu\right|\sqrt{L}
	\label{eq:modulationIndex}
\end{equation}
 is the product of \(\gsquig\), the square root of the efficiency of the IDT \(\left|\tbmu\right|\), and the square root of the length of the interaction region \(L\).  The numerical values for \(\gsquig |\tbmu|\) are overlaid on the measurements in Figure~\ref{fig:Seo}b for the Rayleigh (red) and SH (blue) mode. 

For the measured coupling coefficients in Figure~\ref{fig:AO}b, we extract the peaks of \(h_\trm{oa}\) for each mode and remove a factor of \(|\tbmu|\) determined numerically. The accuracy of the resulting rates are susceptible to errors in \(\tbmu\). In Dahmani~\emph{et al.} where the transmission coefficient was de-embedded directly from measurements, the simulated \(\tbmu\) of 8.9\% was larger than the measured 7.0\% by \(27\%\)~\cite{Dahmani2019}. If we overestimate \(\tbmu\), we will underestimate \(\gsquig\) and therefore the photoelastic coefficients. On the other hand, reflections, \emph{e.g.} off the waveguide, can give rise to standing waves which can enhance the IDT's efficiency. A fractional uncertainty of 27\% like that in Dahmani~\emph{et al.} (plotted in Figure~\ref{fig:AO}b) would not account for the deviation from theory using bulk material properties. We perform a regression on \(\gsquig(\theta_\trm{x})\) in Appendix~\ref{app:regression} to estimate the photoelastic tensor had the deviation been only due to a discrepancy between bulk and our thin-film's $p_{ij}$. We find that scaling $p_{33}, p_{44},$ and $p_{41}$ by factors of 32\%, 70\%, and 35\%, respectively, gives the best fit dashed curve in Figure~\ref{fig:AO}b.

\section{Conclusion}

Lithium niobate-on-sapphire has many bright prospects in optics and, specifically, in acousto-optics. In this platform, the piezoelectric LN film supports both Rayleigh and horizontal shear surface waves which can be generated with interdigital transducers and used to modulate optical waveguides patterned in LN. Here we measure the piezoelectric coupling coefficients of transducers and optomechanical coupling coefficients of ridge waveguides for a range of crystal orientations in X-cut LN, confirming the quality of these bonded films and demonstrating the potential of the material platform. 

As the mechanical frequency reaches into the GHz regime, \(k_\trm{eff}^2\) of the horiztonal shear waves exceeds 10\%, making it possible to make compact, broadband transducers. Future work in pursuit of low-power acousto-optic devices calls for the efficient, mode-selective transduction of wavelength-scale waveguides. Waveguide transducers like those recently developed for horizontal shear waves in suspended LN films can enable a new generation of ultra-low-power phononic devices and acousto-optic modulators~\cite{Dahmani2019}. As the array of electro-optic and nonlinear optical devices in thin-film LN grows, so too grow the prospects for integrating acousto-optic devices into complex phononic and photonic circuits and systems built on these rapidly developing platforms.

\section*{Acknowledgements}

This work was supported by a MURI grant from the U. S. Air Force Office of Scientific Research (Grant No. FA9550-17-1-0002), the DARPA Young Faculty Award (YFA), by a fellowship from the David and Lucille Packard foundation, and by the National Science Foundation through ECCS-1808100 and PHY-1820938. The authors wish to thank NTT Research Inc. for their financial and technical support. Part of this work was performed at the Stanford Nano Shared Facilities (SNSF), supported by the National Science Foundation under Grant No. ECCS-1542152, and the Stanford Nanofabrication Facility (SNF).

\vspace{1em}
The data that support the findings of this study are available from the corresponding author upon reasonable request.

%\bibliography{references}

\begin{thebibliography}{39}%
	\makeatletter
	\providecommand \@ifxundefined [1]{%
		\@ifx{#1\undefined}
	}%
	\providecommand \@ifnum [1]{%
		\ifnum #1\expandafter \@firstoftwo
		\else \expandafter \@secondoftwo
		\fi
	}%
	\providecommand \@ifx [1]{%
		\ifx #1\expandafter \@firstoftwo
		\else \expandafter \@secondoftwo
		\fi
	}%
	\providecommand \natexlab [1]{#1}%
	\providecommand \enquote  [1]{``#1''}%
	\providecommand \bibnamefont  [1]{#1}%
	\providecommand \bibfnamefont [1]{#1}%
	\providecommand \citenamefont [1]{#1}%
	\providecommand \href@noop [0]{\@secondoftwo}%
	\providecommand \href [0]{\begingroup \@sanitize@url \@href}%
	\providecommand \@href[1]{\@@startlink{#1}\@@href}%
	\providecommand \@@href[1]{\endgroup#1\@@endlink}%
	\providecommand \@sanitize@url [0]{\catcode `\\12\catcode `\$12\catcode
		`\&12\catcode `\#12\catcode `\^12\catcode `\_12\catcode `\%12\relax}%
	\providecommand \@@startlink[1]{}%
	\providecommand \@@endlink[0]{}%
	\providecommand \url  [0]{\begingroup\@sanitize@url \@url }%
	\providecommand \@url [1]{\endgroup\@href {#1}{\urlprefix }}%
	\providecommand \urlprefix  [0]{URL }%
	\providecommand \Eprint [0]{\href }%
	\providecommand \doibase [0]{http://dx.doi.org/}%
	\providecommand \selectlanguage [0]{\@gobble}%
	\providecommand \bibinfo  [0]{\@secondoftwo}%
	\providecommand \bibfield  [0]{\@secondoftwo}%
	\providecommand \translation [1]{[#1]}%
	\providecommand \BibitemOpen [0]{}%
	\providecommand \bibitemStop [0]{}%
	\providecommand \bibitemNoStop [0]{.\EOS\space}%
	\providecommand \EOS [0]{\spacefactor3000\relax}%
	\providecommand \BibitemShut  [1]{\csname bibitem#1\endcsname}%
	\let\auto@bib@innerbib\@empty
	%</preamble>
	\bibitem [{\citenamefont {Wang}\ \emph {et~al.}(2014)\citenamefont {Wang},
		\citenamefont {Burek}, \citenamefont {Lin}, \citenamefont {Atikian},
		\citenamefont {Venkataraman}, \citenamefont {Huang}, \citenamefont {Stark},\
		and\ \citenamefont {Lon{\v{c}}ar}}]{Wang2014}%
	\BibitemOpen
	\bibfield  {author} {\bibinfo {author} {\bibfnamefont {C.}~\bibnamefont
			{Wang}}, \bibinfo {author} {\bibfnamefont {M.~J.}\ \bibnamefont {Burek}},
		\bibinfo {author} {\bibfnamefont {Z.}~\bibnamefont {Lin}}, \bibinfo {author}
		{\bibfnamefont {H.~A.}\ \bibnamefont {Atikian}}, \bibinfo {author}
		{\bibfnamefont {V.}~\bibnamefont {Venkataraman}}, \bibinfo {author}
		{\bibfnamefont {I.-C.}\ \bibnamefont {Huang}}, \bibinfo {author}
		{\bibfnamefont {P.}~\bibnamefont {Stark}}, \ and\ \bibinfo {author}
		{\bibfnamefont {M.}~\bibnamefont {Lon{\v{c}}ar}},\ }\bibfield  {title}
	{\enquote {\bibinfo {title} {Integrated high quality factor lithium niobate
				microdisk resonators},}\ }\href@noop {} {\bibfield  {journal} {\bibinfo
			{journal} {Optics express}\ }\textbf {\bibinfo {volume} {22}},\ \bibinfo
		{pages} {30924--30933} (\bibinfo {year} {2014})}\BibitemShut {NoStop}%
	\bibitem [{\citenamefont {Levy}\ \emph {et~al.}(1998)\citenamefont {Levy},
		\citenamefont {Osgood~Jr}, \citenamefont {Liu}, \citenamefont {Cross},
		\citenamefont {Cargill~III}, \citenamefont {Kumar},\ and\ \citenamefont
		{Bakhru}}]{Levy1998}%
	\BibitemOpen
	\bibfield  {author} {\bibinfo {author} {\bibfnamefont {M.}~\bibnamefont
			{Levy}}, \bibinfo {author} {\bibfnamefont {R.}~\bibnamefont {Osgood~Jr}},
		\bibinfo {author} {\bibfnamefont {R.}~\bibnamefont {Liu}}, \bibinfo {author}
		{\bibfnamefont {L.}~\bibnamefont {Cross}}, \bibinfo {author} {\bibfnamefont
			{G.}~\bibnamefont {Cargill~III}}, \bibinfo {author} {\bibfnamefont
			{A.}~\bibnamefont {Kumar}}, \ and\ \bibinfo {author} {\bibfnamefont
			{H.}~\bibnamefont {Bakhru}},\ }\bibfield  {title} {\enquote {\bibinfo {title}
			{Fabrication of single-crystal lithium niobate films by crystal ion
				slicing},}\ }\href@noop {} {\bibfield  {journal} {\bibinfo  {journal}
			{Applied Physics Letters}\ }\textbf {\bibinfo {volume} {73}},\ \bibinfo
		{pages} {2293--2295} (\bibinfo {year} {1998})}\BibitemShut {NoStop}%
	\bibitem [{\citenamefont {Yu}\ \emph {et~al.}(2019)\citenamefont {Yu},
		\citenamefont {Wang}, \citenamefont {Zhang},\ and\ \citenamefont
		{Loncar}}]{Yu2019}%
	\BibitemOpen
	\bibfield  {author} {\bibinfo {author} {\bibfnamefont {M.}~\bibnamefont
			{Yu}}, \bibinfo {author} {\bibfnamefont {C.}~\bibnamefont {Wang}}, \bibinfo
		{author} {\bibfnamefont {M.}~\bibnamefont {Zhang}}, \ and\ \bibinfo {author}
		{\bibfnamefont {M.}~\bibnamefont {Loncar}},\ }\bibfield  {title} {\enquote
		{\bibinfo {title} {Chip-based lithium-niobate frequency combs},}\ }\href@noop
	{} {\bibfield  {journal} {\bibinfo  {journal} {IEEE Photonics Technology
				Letters}\ } (\bibinfo {year} {2019})}\BibitemShut {NoStop}%
	\bibitem [{\citenamefont {Rao}\ \emph {et~al.}(2016{\natexlab{a}})\citenamefont
		{Rao}, \citenamefont {Malinowski}, \citenamefont {Honardoost}, \citenamefont
		{Talukder}, \citenamefont {Rabiei}, \citenamefont {Delfyett},\ and\
		\citenamefont {Fathpour}}]{Rao2016}%
	\BibitemOpen
	\bibfield  {author} {\bibinfo {author} {\bibfnamefont {A.}~\bibnamefont
			{Rao}}, \bibinfo {author} {\bibfnamefont {M.}~\bibnamefont {Malinowski}},
		\bibinfo {author} {\bibfnamefont {A.}~\bibnamefont {Honardoost}}, \bibinfo
		{author} {\bibfnamefont {J.~R.}\ \bibnamefont {Talukder}}, \bibinfo {author}
		{\bibfnamefont {P.}~\bibnamefont {Rabiei}}, \bibinfo {author} {\bibfnamefont
			{P.}~\bibnamefont {Delfyett}}, \ and\ \bibinfo {author} {\bibfnamefont
			{S.}~\bibnamefont {Fathpour}},\ }\bibfield  {title} {\enquote {\bibinfo
			{title} {Second-harmonic generation in periodically-poled thin film lithium
				niobate wafer-bonded on silicon},}\ }\href@noop {} {\bibfield  {journal}
		{\bibinfo  {journal} {Optics express}\ }\textbf {\bibinfo {volume} {24}},\
		\bibinfo {pages} {29941--29947} (\bibinfo {year}
		{2016}{\natexlab{a}})}\BibitemShut {NoStop}%
	\bibitem [{\citenamefont {Chang}\ \emph {et~al.}(2016)\citenamefont {Chang},
		\citenamefont {Li}, \citenamefont {Volet}, \citenamefont {Wang},
		\citenamefont {Peters},\ and\ \citenamefont {Bowers}}]{Chang2016}%
	\BibitemOpen
	\bibfield  {author} {\bibinfo {author} {\bibfnamefont {L.}~\bibnamefont
			{Chang}}, \bibinfo {author} {\bibfnamefont {Y.}~\bibnamefont {Li}}, \bibinfo
		{author} {\bibfnamefont {N.}~\bibnamefont {Volet}}, \bibinfo {author}
		{\bibfnamefont {L.}~\bibnamefont {Wang}}, \bibinfo {author} {\bibfnamefont
			{J.}~\bibnamefont {Peters}}, \ and\ \bibinfo {author} {\bibfnamefont {J.~E.}\
			\bibnamefont {Bowers}},\ }\bibfield  {title} {\enquote {\bibinfo {title}
			{Thin film wavelength converters for photonic integrated circuits},}\
	}\href@noop {} {\bibfield  {journal} {\bibinfo  {journal} {Optica}\ }\textbf
		{\bibinfo {volume} {3}},\ \bibinfo {pages} {531--535} (\bibinfo {year}
		{2016})}\BibitemShut {NoStop}%
	\bibitem [{\citenamefont {Rao}\ \emph {et~al.}(2016{\natexlab{b}})\citenamefont
		{Rao}, \citenamefont {Patil}, \citenamefont {Rabiei}, \citenamefont
		{Honardoost}, \citenamefont {DeSalvo}, \citenamefont {Paolella},\ and\
		\citenamefont {Fathpour}}]{Rao2016b}%
	\BibitemOpen
	\bibfield  {author} {\bibinfo {author} {\bibfnamefont {A.}~\bibnamefont
			{Rao}}, \bibinfo {author} {\bibfnamefont {A.}~\bibnamefont {Patil}}, \bibinfo
		{author} {\bibfnamefont {P.}~\bibnamefont {Rabiei}}, \bibinfo {author}
		{\bibfnamefont {A.}~\bibnamefont {Honardoost}}, \bibinfo {author}
		{\bibfnamefont {R.}~\bibnamefont {DeSalvo}}, \bibinfo {author} {\bibfnamefont
			{A.}~\bibnamefont {Paolella}}, \ and\ \bibinfo {author} {\bibfnamefont
			{S.}~\bibnamefont {Fathpour}},\ }\bibfield  {title} {\enquote {\bibinfo
			{title} {High-performance and linear thin-film lithium niobate mach--zehnder
				modulators on silicon up to 50 ghz},}\ }\href@noop {} {\bibfield  {journal}
		{\bibinfo  {journal} {Optics letters}\ }\textbf {\bibinfo {volume} {41}},\
		\bibinfo {pages} {5700--5703} (\bibinfo {year}
		{2016}{\natexlab{b}})}\BibitemShut {NoStop}%
	\bibitem [{\citenamefont {Wang}\ \emph {et~al.}(2018)\citenamefont {Wang},
		\citenamefont {Zhang}, \citenamefont {Stern}, \citenamefont {Lipson},\ and\
		\citenamefont {Lon{\v{c}}ar}}]{Wang2018}%
	\BibitemOpen
	\bibfield  {author} {\bibinfo {author} {\bibfnamefont {C.}~\bibnamefont
			{Wang}}, \bibinfo {author} {\bibfnamefont {M.}~\bibnamefont {Zhang}},
		\bibinfo {author} {\bibfnamefont {B.}~\bibnamefont {Stern}}, \bibinfo
		{author} {\bibfnamefont {M.}~\bibnamefont {Lipson}}, \ and\ \bibinfo {author}
		{\bibfnamefont {M.}~\bibnamefont {Lon{\v{c}}ar}},\ }\bibfield  {title}
	{\enquote {\bibinfo {title} {Nanophotonic lithium niobate electro-optic
				modulators},}\ }\href@noop {} {\bibfield  {journal} {\bibinfo  {journal}
			{Optics express}\ }\textbf {\bibinfo {volume} {26}},\ \bibinfo {pages}
		{1547--1555} (\bibinfo {year} {2018})}\BibitemShut {NoStop}%
	\bibitem [{\citenamefont {Olsson~III}\ \emph {et~al.}(2014)\citenamefont
		{Olsson~III}, \citenamefont {Hattar}, \citenamefont {Homeijer}, \citenamefont
		{Wiwi}, \citenamefont {Eichenfield}, \citenamefont {Branch}, \citenamefont
		{Baker}, \citenamefont {Nguyen}, \citenamefont {Clark}, \citenamefont
		{Bauer},\ and\ \citenamefont {Friedmann}}]{Olsson2014}%
	\BibitemOpen
	\bibfield  {author} {\bibinfo {author} {\bibfnamefont {R.~H.}\ \bibnamefont
			{Olsson~III}}, \bibinfo {author} {\bibfnamefont {K.}~\bibnamefont {Hattar}},
		\bibinfo {author} {\bibfnamefont {S.~J.}\ \bibnamefont {Homeijer}}, \bibinfo
		{author} {\bibfnamefont {M.}~\bibnamefont {Wiwi}}, \bibinfo {author}
		{\bibfnamefont {M.}~\bibnamefont {Eichenfield}}, \bibinfo {author}
		{\bibfnamefont {D.~W.}\ \bibnamefont {Branch}}, \bibinfo {author}
		{\bibfnamefont {M.~S.}\ \bibnamefont {Baker}}, \bibinfo {author}
		{\bibfnamefont {J.}~\bibnamefont {Nguyen}}, \bibinfo {author} {\bibfnamefont
			{B.}~\bibnamefont {Clark}}, \bibinfo {author} {\bibfnamefont
			{T.}~\bibnamefont {Bauer}}, \ and\ \bibinfo {author} {\bibfnamefont {T.~A.}\
			\bibnamefont {Friedmann}},\ }\bibfield  {title} {\enquote {\bibinfo {title}
			{A high electromechanical coupling coefficient sh0 lamb wave lithium niobate
				micromechanical resonator and a method for fabrication},}\ }\href@noop {}
	{\bibfield  {journal} {\bibinfo  {journal} {Sensors and Actuators A:
				Physical}\ }\textbf {\bibinfo {volume} {209}},\ \bibinfo {pages} {183--190}
		(\bibinfo {year} {2014})}\BibitemShut {NoStop}%
	\bibitem [{\citenamefont {Pop}\ \emph {et~al.}(2017)\citenamefont {Pop},
		\citenamefont {Kochhar}, \citenamefont {Vidal-Alvarez},\ and\ \citenamefont
		{Piazza}}]{Pop2017}%
	\BibitemOpen
	\bibfield  {author} {\bibinfo {author} {\bibfnamefont {F.~V.}\ \bibnamefont
			{Pop}}, \bibinfo {author} {\bibfnamefont {A.~S.}\ \bibnamefont {Kochhar}},
		\bibinfo {author} {\bibfnamefont {G.}~\bibnamefont {Vidal-Alvarez}}, \ and\
		\bibinfo {author} {\bibfnamefont {G.}~\bibnamefont {Piazza}},\ }\bibfield
	{title} {\enquote {\bibinfo {title} {Laterally vibrating lithium niobate mems
				resonators with 30\% electromechanical coupling coefficient},}\ }in\
	\href@noop {} {\emph {\bibinfo {booktitle} {2017 IEEE 30th International
				Conference on Micro Electro Mechanical Systems (MEMS)}}}\ (\bibinfo
	{organization} {IEEE},\ \bibinfo {year} {2017})\ pp.\ \bibinfo {pages}
	{966--969}\BibitemShut {NoStop}%
	\bibitem [{\citenamefont {Manzaneque}\ \emph {et~al.}(2019)\citenamefont
		{Manzaneque}, \citenamefont {Lu}, \citenamefont {Yang},\ and\ \citenamefont
		{Gong}}]{Manzaneque2019}%
	\BibitemOpen
	\bibfield  {author} {\bibinfo {author} {\bibfnamefont {T.}~\bibnamefont
			{Manzaneque}}, \bibinfo {author} {\bibfnamefont {R.}~\bibnamefont {Lu}},
		\bibinfo {author} {\bibfnamefont {Y.}~\bibnamefont {Yang}}, \ and\ \bibinfo
		{author} {\bibfnamefont {S.}~\bibnamefont {Gong}},\ }\bibfield  {title}
	{\enquote {\bibinfo {title} {Low-loss and wideband acoustic delay lines},}\
	}\href@noop {} {\bibfield  {journal} {\bibinfo  {journal} {IEEE Transactions
				on Microwave Theory and Techniques}\ }\textbf {\bibinfo {volume} {67}},\
		\bibinfo {pages} {1379--1391} (\bibinfo {year} {2019})}\BibitemShut {NoStop}%
	\bibitem [{\citenamefont {Dahmani}\ \emph {et~al.}(2019)\citenamefont
		{Dahmani}, \citenamefont {Sarabalis}, \citenamefont {Jiang}, \citenamefont
		{Mayor},\ and\ \citenamefont {Safavi-Naeini}}]{Dahmani2019}%
	\BibitemOpen
	\bibfield  {author} {\bibinfo {author} {\bibfnamefont {Y.~D.}\ \bibnamefont
			{Dahmani}}, \bibinfo {author} {\bibfnamefont {C.~J.}\ \bibnamefont
			{Sarabalis}}, \bibinfo {author} {\bibfnamefont {W.}~\bibnamefont {Jiang}},
		\bibinfo {author} {\bibfnamefont {F.~M.}\ \bibnamefont {Mayor}}, \ and\
		\bibinfo {author} {\bibfnamefont {A.~H.}\ \bibnamefont {Safavi-Naeini}},\
	}\bibfield  {title} {\enquote {\bibinfo {title} {Piezoelectric transduction
				of a wavelength-scale mechanical waveguide},}\ }\href@noop {} {\bibfield
		{journal} {\bibinfo  {journal} {arXiv preprint arXiv:1907.13058}\ } (\bibinfo
		{year} {2019})}\BibitemShut {NoStop}%
	\bibitem [{\citenamefont {Shao}\ \emph {et~al.}(2019)\citenamefont {Shao},
		\citenamefont {Yu}, \citenamefont {Maity}, \citenamefont {Sinclair},
		\citenamefont {Zheng}, \citenamefont {Chia}, \citenamefont {Shams-Ansari},
		\citenamefont {Wang}, \citenamefont {Zhang}, \citenamefont {Lai},\ and\
		\citenamefont {Lon\v{c}ar}}]{Shao2019}%
	\BibitemOpen
	\bibfield  {author} {\bibinfo {author} {\bibfnamefont {L.}~\bibnamefont
			{Shao}}, \bibinfo {author} {\bibfnamefont {M.}~\bibnamefont {Yu}}, \bibinfo
		{author} {\bibfnamefont {S.}~\bibnamefont {Maity}}, \bibinfo {author}
		{\bibfnamefont {N.}~\bibnamefont {Sinclair}}, \bibinfo {author}
		{\bibfnamefont {L.}~\bibnamefont {Zheng}}, \bibinfo {author} {\bibfnamefont
			{C.}~\bibnamefont {Chia}}, \bibinfo {author} {\bibfnamefont {A.}~\bibnamefont
			{Shams-Ansari}}, \bibinfo {author} {\bibfnamefont {C.}~\bibnamefont {Wang}},
		\bibinfo {author} {\bibfnamefont {M.}~\bibnamefont {Zhang}}, \bibinfo
		{author} {\bibfnamefont {K.}~\bibnamefont {Lai}}, \ and\ \bibinfo {author}
		{\bibfnamefont {M.}~\bibnamefont {Lon\v{c}ar}},\ }\bibfield  {title}
	{\enquote {\bibinfo {title} {Microwave-to-optical conversion using lithium
				niobate thin-film acoustic resonators},}\ }\href@noop {} {\bibfield
		{journal} {\bibinfo  {journal} {Optica}\ }\textbf {\bibinfo {volume} {6}},\
		\bibinfo {pages} {1498--1505} (\bibinfo {year} {2019})}\BibitemShut {NoStop}%
	\bibitem [{\citenamefont {Jiang}\ \emph {et~al.}(2020)\citenamefont {Jiang},
		\citenamefont {Sarabalis}, \citenamefont {Dahmani}, \citenamefont {Patel},
		\citenamefont {Mayor}, \citenamefont {McKenna}, \citenamefont {Van~Laer},\
		and\ \citenamefont {Safavi-Naeini}}]{Jiang2019}%
	\BibitemOpen
	\bibfield  {author} {\bibinfo {author} {\bibfnamefont {W.}~\bibnamefont
			{Jiang}}, \bibinfo {author} {\bibfnamefont {C.~J.}\ \bibnamefont
			{Sarabalis}}, \bibinfo {author} {\bibfnamefont {Y.~D.}\ \bibnamefont
			{Dahmani}}, \bibinfo {author} {\bibfnamefont {R.~N.}\ \bibnamefont {Patel}},
		\bibinfo {author} {\bibfnamefont {F.~M.}\ \bibnamefont {Mayor}}, \bibinfo
		{author} {\bibfnamefont {T.~P.}\ \bibnamefont {McKenna}}, \bibinfo {author}
		{\bibfnamefont {R.}~\bibnamefont {Van~Laer}}, \ and\ \bibinfo {author}
		{\bibfnamefont {A.~H.}\ \bibnamefont {Safavi-Naeini}},\ }\bibfield  {title}
	{\enquote {\bibinfo {title} {Efficient bidirectional piezo-optomechanical
				transduction between microwave and optical frequency},}\ }\href@noop {}
	{\bibfield  {journal} {\bibinfo  {journal} {Nature Communications}\ }\textbf
		{\bibinfo {volume} {11}},\ \bibinfo {pages} {1--7} (\bibinfo {year}
		{2020})}\BibitemShut {NoStop}%
	\bibitem [{\citenamefont {Tadesse}\ and\ \citenamefont
		{Li}(2014)}]{Tadesse2014}%
	\BibitemOpen
	\bibfield  {author} {\bibinfo {author} {\bibfnamefont {S.~A.}\ \bibnamefont
			{Tadesse}}\ and\ \bibinfo {author} {\bibfnamefont {M.}~\bibnamefont {Li}},\
	}\bibfield  {title} {\enquote {\bibinfo {title} {{Sub-optical wavelength
					acoustic wave modulation of integrated photonic resonators at microwave
					frequencies}},}\ }\href {\doibase 10.1038/ncomms6402} {\bibfield  {journal}
		{\bibinfo  {journal} {Nature Communications}\ }\textbf {\bibinfo {volume}
			{5}},\ \bibinfo {pages} {5402} (\bibinfo {year} {2014})}\BibitemShut
	{NoStop}%
	\bibitem [{\citenamefont {Kittlaus}\ \emph {et~al.}(2020)\citenamefont
		{Kittlaus}, \citenamefont {Jones}, \citenamefont {Rakich}, \citenamefont
		{Otterstrom}, \citenamefont {Muller},\ and\ \citenamefont
		{Rais-Zadeh}}]{Kittlaus2020}%
	\BibitemOpen
	\bibfield  {author} {\bibinfo {author} {\bibfnamefont {E.~A.}\ \bibnamefont
			{Kittlaus}}, \bibinfo {author} {\bibfnamefont {W.~M.}\ \bibnamefont {Jones}},
		\bibinfo {author} {\bibfnamefont {P.~T.}\ \bibnamefont {Rakich}}, \bibinfo
		{author} {\bibfnamefont {N.~T.}\ \bibnamefont {Otterstrom}}, \bibinfo
		{author} {\bibfnamefont {R.~E.}\ \bibnamefont {Muller}}, \ and\ \bibinfo
		{author} {\bibfnamefont {M.}~\bibnamefont {Rais-Zadeh}},\ }\bibfield  {title}
	{\enquote {\bibinfo {title} {Electrically-driven acousto-optics and broadband
				non-reciprocity in silicon photonics},}\ }\href@noop {} {\bibfield  {journal}
		{\bibinfo  {journal} {arXiv preprint arXiv:2004.01270}\ } (\bibinfo {year}
		{2020})}\BibitemShut {NoStop}%
	\bibitem [{\citenamefont {van~der Poel}\ \emph {et~al.}(2007)\citenamefont
		{van~der Poel}, \citenamefont {Beck}, \citenamefont {D{\"u}hring},
		\citenamefont {de~Lima}, \citenamefont {Frandsen}, \citenamefont {Peucheret},
		\citenamefont {Sigmund}, \citenamefont {Jahn}, \citenamefont {Hvam},\ and\
		\citenamefont {Santos}}]{VanDerPoel2007}%
	\BibitemOpen
	\bibfield  {author} {\bibinfo {author} {\bibfnamefont {M.}~\bibnamefont
			{van~der Poel}}, \bibinfo {author} {\bibfnamefont {M.}~\bibnamefont {Beck}},
		\bibinfo {author} {\bibfnamefont {M.~B.}\ \bibnamefont {D{\"u}hring}},
		\bibinfo {author} {\bibfnamefont {M.~M.}\ \bibnamefont {de~Lima}}, \bibinfo
		{author} {\bibfnamefont {L.~H.}\ \bibnamefont {Frandsen}}, \bibinfo {author}
		{\bibfnamefont {C.}~\bibnamefont {Peucheret}}, \bibinfo {author}
		{\bibfnamefont {O.}~\bibnamefont {Sigmund}}, \bibinfo {author} {\bibfnamefont
			{U.}~\bibnamefont {Jahn}}, \bibinfo {author} {\bibfnamefont {J.~M.}\
			\bibnamefont {Hvam}}, \ and\ \bibinfo {author} {\bibfnamefont
			{P.}~\bibnamefont {Santos}},\ }\bibfield  {title} {\enquote {\bibinfo {title}
			{Surface acoustic wave driven light modulation},}\ }in\ \href@noop {} {\emph
		{\bibinfo {booktitle} {Proceedings of 13th European Conference on Integrated
				Optics}}}\ (\bibinfo {organization} {Citeseer},\ \bibinfo {year} {2007})\ p.\
	\bibinfo {pages} {FB3}\BibitemShut {NoStop}%
	\bibitem [{\citenamefont {Van Der~Slot}, \citenamefont {Porcel},\ and\
		\citenamefont {Boller}(2019)}]{Slot2019}%
	\BibitemOpen
	\bibfield  {author} {\bibinfo {author} {\bibfnamefont {P.~J.}\ \bibnamefont
			{Van Der~Slot}}, \bibinfo {author} {\bibfnamefont {M.~A.}\ \bibnamefont
			{Porcel}}, \ and\ \bibinfo {author} {\bibfnamefont {K.-J.}\ \bibnamefont
			{Boller}},\ }\bibfield  {title} {\enquote {\bibinfo {title} {Surface acoustic
				waves for acousto-optic modulation in buried silicon nitride waveguides},}\
	}\href@noop {} {\bibfield  {journal} {\bibinfo  {journal} {Optics express}\
		}\textbf {\bibinfo {volume} {27}},\ \bibinfo {pages} {1433--1452} (\bibinfo
		{year} {2019})}\BibitemShut {NoStop}%
	\bibitem [{\citenamefont {Cai}\ \emph {et~al.}(2019)\citenamefont {Cai},
		\citenamefont {Mahmoud}, \citenamefont {Khan}, \citenamefont {Mahmoud},
		\citenamefont {Mukherjee}, \citenamefont {Bain},\ and\ \citenamefont
		{Piazza}}]{Cai2019}%
	\BibitemOpen
	\bibfield  {author} {\bibinfo {author} {\bibfnamefont {L.}~\bibnamefont
			{Cai}}, \bibinfo {author} {\bibfnamefont {A.}~\bibnamefont {Mahmoud}},
		\bibinfo {author} {\bibfnamefont {M.}~\bibnamefont {Khan}}, \bibinfo {author}
		{\bibfnamefont {M.}~\bibnamefont {Mahmoud}}, \bibinfo {author} {\bibfnamefont
			{T.}~\bibnamefont {Mukherjee}}, \bibinfo {author} {\bibfnamefont
			{J.}~\bibnamefont {Bain}}, \ and\ \bibinfo {author} {\bibfnamefont
			{G.}~\bibnamefont {Piazza}},\ }\bibfield  {title} {\enquote {\bibinfo {title}
			{Acousto-optical modulation of thin film lithium niobate waveguide
				devices},}\ }\href@noop {} {\bibfield  {journal} {\bibinfo  {journal}
			{Photonics Research}\ }\textbf {\bibinfo {volume} {7}},\ \bibinfo {pages}
		{1003--1013} (\bibinfo {year} {2019})}\BibitemShut {NoStop}%
	\bibitem [{\citenamefont {Mahmoud}\ \emph
		{et~al.}(2018{\natexlab{a}})\citenamefont {Mahmoud}, \citenamefont {Mahmoud},
		\citenamefont {Cai}, \citenamefont {Khan}, \citenamefont {Mukherjee},
		\citenamefont {Bain},\ and\ \citenamefont {Piazza}}]{Mahmoud2018}%
	\BibitemOpen
	\bibfield  {author} {\bibinfo {author} {\bibfnamefont {M.}~\bibnamefont
			{Mahmoud}}, \bibinfo {author} {\bibfnamefont {A.}~\bibnamefont {Mahmoud}},
		\bibinfo {author} {\bibfnamefont {L.}~\bibnamefont {Cai}}, \bibinfo {author}
		{\bibfnamefont {M.}~\bibnamefont {Khan}}, \bibinfo {author} {\bibfnamefont
			{T.}~\bibnamefont {Mukherjee}}, \bibinfo {author} {\bibfnamefont
			{J.}~\bibnamefont {Bain}}, \ and\ \bibinfo {author} {\bibfnamefont
			{G.}~\bibnamefont {Piazza}},\ }\bibfield  {title} {\enquote {\bibinfo {title}
			{Novel on chip rotation detection based on the acousto-optic effect in
				surface acoustic wave gyroscopes},}\ }\href@noop {} {\bibfield  {journal}
		{\bibinfo  {journal} {Optics express}\ }\textbf {\bibinfo {volume} {26}},\
		\bibinfo {pages} {25060--25075} (\bibinfo {year}
		{2018}{\natexlab{a}})}\BibitemShut {NoStop}%
	\bibitem [{\citenamefont {Mahmoud}\ \emph
		{et~al.}(2018{\natexlab{b}})\citenamefont {Mahmoud}, \citenamefont {Mahmoud},
		\citenamefont {Cai}, \citenamefont {Khan}, \citenamefont {Bain},
		\citenamefont {Mukherjee},\ and\ \citenamefont {Piazza}}]{Mahmoud2018b}%
	\BibitemOpen
	\bibfield  {author} {\bibinfo {author} {\bibfnamefont {A.}~\bibnamefont
			{Mahmoud}}, \bibinfo {author} {\bibfnamefont {M.}~\bibnamefont {Mahmoud}},
		\bibinfo {author} {\bibfnamefont {L.}~\bibnamefont {Cai}}, \bibinfo {author}
		{\bibfnamefont {M.}~\bibnamefont {Khan}}, \bibinfo {author} {\bibfnamefont
			{J.~A.}\ \bibnamefont {Bain}}, \bibinfo {author} {\bibfnamefont
			{T.}~\bibnamefont {Mukherjee}}, \ and\ \bibinfo {author} {\bibfnamefont
			{G.}~\bibnamefont {Piazza}},\ }\bibfield  {title} {\enquote {\bibinfo {title}
			{Acousto-optic gyroscope},}\ }in\ \href@noop {} {\emph {\bibinfo {booktitle}
			{2018 IEEE Micro Electro Mechanical Systems (MEMS)}}}\ (\bibinfo
	{organization} {IEEE},\ \bibinfo {year} {2018})\ pp.\ \bibinfo {pages}
	{241--244}\BibitemShut {NoStop}%
	\bibitem [{\citenamefont {Crespo-Poveda}\ \emph {et~al.}(2015)\citenamefont
		{Crespo-Poveda}, \citenamefont {Hern{\'a}ndez-M{\'\i}nguez}, \citenamefont
		{Gargallo}, \citenamefont {Biermann}, \citenamefont {Tahraoui}, \citenamefont
		{Santos}, \citenamefont {Mu{\~n}oz}, \citenamefont {Cantarero},\ and\
		\citenamefont {de~Lima}}]{Poveda2015}%
	\BibitemOpen
	\bibfield  {author} {\bibinfo {author} {\bibfnamefont {A.}~\bibnamefont
			{Crespo-Poveda}}, \bibinfo {author} {\bibfnamefont {A.}~\bibnamefont
			{Hern{\'a}ndez-M{\'\i}nguez}}, \bibinfo {author} {\bibfnamefont
			{B.}~\bibnamefont {Gargallo}}, \bibinfo {author} {\bibfnamefont
			{K.}~\bibnamefont {Biermann}}, \bibinfo {author} {\bibfnamefont
			{A.}~\bibnamefont {Tahraoui}}, \bibinfo {author} {\bibfnamefont
			{P.}~\bibnamefont {Santos}}, \bibinfo {author} {\bibfnamefont
			{P.}~\bibnamefont {Mu{\~n}oz}}, \bibinfo {author} {\bibfnamefont
			{A.}~\bibnamefont {Cantarero}}, \ and\ \bibinfo {author} {\bibfnamefont
			{M.}~\bibnamefont {de~Lima}},\ }\bibfield  {title} {\enquote {\bibinfo
			{title} {Acoustically driven arrayed waveguide grating},}\ }\href@noop {}
	{\bibfield  {journal} {\bibinfo  {journal} {Optics express}\ }\textbf
		{\bibinfo {volume} {23}},\ \bibinfo {pages} {21213--21231} (\bibinfo {year}
		{2015})}\BibitemShut {NoStop}%
	\bibitem [{\citenamefont {Crespo-Poveda}\ \emph {et~al.}(2016)\citenamefont
		{Crespo-Poveda}, \citenamefont {Hern{\'a}ndez-M{\'\i}nguez}, \citenamefont
		{Biermann}, \citenamefont {Tahraoui}, \citenamefont {Gargallo}, \citenamefont
		{Mu{\~n}oz}, \citenamefont {Santos}, \citenamefont {Cantarero},\ and\
		\citenamefont {de~Lima~Jr}}]{Poveda2016}%
	\BibitemOpen
	\bibfield  {author} {\bibinfo {author} {\bibfnamefont {A.}~\bibnamefont
			{Crespo-Poveda}}, \bibinfo {author} {\bibfnamefont {A.}~\bibnamefont
			{Hern{\'a}ndez-M{\'\i}nguez}}, \bibinfo {author} {\bibfnamefont
			{K.}~\bibnamefont {Biermann}}, \bibinfo {author} {\bibfnamefont
			{A.}~\bibnamefont {Tahraoui}}, \bibinfo {author} {\bibfnamefont
			{B.}~\bibnamefont {Gargallo}}, \bibinfo {author} {\bibfnamefont
			{P.}~\bibnamefont {Mu{\~n}oz}}, \bibinfo {author} {\bibfnamefont {P.~V.}\
			\bibnamefont {Santos}}, \bibinfo {author} {\bibfnamefont {A.}~\bibnamefont
			{Cantarero}}, \ and\ \bibinfo {author} {\bibfnamefont {M.~M.}\ \bibnamefont
			{de~Lima~Jr}},\ }\bibfield  {title} {\enquote {\bibinfo {title} {Tunable
				arrayed waveguide grating driven by surface acoustic waves},}\ }in\
	\href@noop {} {\emph {\bibinfo {booktitle} {Smart Photonic and Optoelectronic
				Integrated Circuits XVIII}}},\ Vol.\ \bibinfo {volume} {9751}\ (\bibinfo
	{organization} {International Society for Optics and Photonics},\ \bibinfo
	{year} {2016})\ p.\ \bibinfo {pages} {97510Y}\BibitemShut {NoStop}%
	\bibitem [{\citenamefont {Khan}\ \emph {et~al.}(2019)\citenamefont {Khan},
		\citenamefont {Mahmoud}, \citenamefont {Cai}, \citenamefont {Mahmoud},
		\citenamefont {Mukherjee}, \citenamefont {Bain},\ and\ \citenamefont
		{Piazza}}]{Khan2019}%
	\BibitemOpen
	\bibfield  {author} {\bibinfo {author} {\bibfnamefont {M.}~\bibnamefont
			{Khan}}, \bibinfo {author} {\bibfnamefont {A.}~\bibnamefont {Mahmoud}},
		\bibinfo {author} {\bibfnamefont {L.}~\bibnamefont {Cai}}, \bibinfo {author}
		{\bibfnamefont {M.}~\bibnamefont {Mahmoud}}, \bibinfo {author} {\bibfnamefont
			{T.}~\bibnamefont {Mukherjee}}, \bibinfo {author} {\bibfnamefont {J.~A.}\
			\bibnamefont {Bain}}, \ and\ \bibinfo {author} {\bibfnamefont
			{G.}~\bibnamefont {Piazza}},\ }\bibfield  {title} {\enquote {\bibinfo {title}
			{Extraction of elasto-optic coefficient of thin film arsenic trisulfide using
				a mach-zehnder acousto-optic modulator on lithium niobate},}\ }\href@noop {}
	{\bibfield  {journal} {\bibinfo  {journal} {Journal of Lightwave Technology}\
		} (\bibinfo {year} {2019})}\BibitemShut {NoStop}%
	\bibitem [{\citenamefont {Hashimoto}\ and\ \citenamefont
		{Hashimoto}(2000)}]{Hashimoto2000}%
	\BibitemOpen
	\bibfield  {author} {\bibinfo {author} {\bibfnamefont {K.-y.}\ \bibnamefont
			{Hashimoto}}\ and\ \bibinfo {author} {\bibfnamefont {K.-Y.}\ \bibnamefont
			{Hashimoto}},\ }\href@noop {} {\emph {\bibinfo {title} {Surface acoustic wave
				devices in telecommunications}}}\ (\bibinfo  {publisher} {Springer},\
	\bibinfo {year} {2000})\BibitemShut {NoStop}%
	\bibitem [{\citenamefont {Hashimoto}(2009)}]{Hashimoto2009}%
	\BibitemOpen
	\bibfield  {author} {\bibinfo {author} {\bibfnamefont {K.-y.}\ \bibnamefont
			{Hashimoto}},\ }\href@noop {} {\emph {\bibinfo {title} {RF bulk acoustic wave
				filters for communications}}}\ (\bibinfo  {publisher} {Artech House},\
	\bibinfo {year} {2009})\BibitemShut {NoStop}%
	\bibitem [{\citenamefont {Andrushchak}\ \emph {et~al.}(2009)\citenamefont
		{Andrushchak}, \citenamefont {Mytsyk}, \citenamefont {Laba}, \citenamefont
		{Yurkevych}, \citenamefont {Solskii}, \citenamefont {Kityk},\ and\
		\citenamefont {Sahraoui}}]{Andrushchak2009}%
	\BibitemOpen
	\bibfield  {author} {\bibinfo {author} {\bibfnamefont {A.}~\bibnamefont
			{Andrushchak}}, \bibinfo {author} {\bibfnamefont {B.}~\bibnamefont {Mytsyk}},
		\bibinfo {author} {\bibfnamefont {H.}~\bibnamefont {Laba}}, \bibinfo {author}
		{\bibfnamefont {O.}~\bibnamefont {Yurkevych}}, \bibinfo {author}
		{\bibfnamefont {I.}~\bibnamefont {Solskii}}, \bibinfo {author} {\bibfnamefont
			{A.}~\bibnamefont {Kityk}}, \ and\ \bibinfo {author} {\bibfnamefont
			{B.}~\bibnamefont {Sahraoui}},\ }\bibfield  {title} {\enquote {\bibinfo
			{title} {Complete sets of elastic constants and photoelastic coefficients of
				pure and mgo-doped lithium niobate crystals at room temperature},}\
	}\href@noop {} {\bibfield  {journal} {\bibinfo  {journal} {Journal of Applied
				Physics}\ }\textbf {\bibinfo {volume} {106}},\ \bibinfo {pages} {073510}
		(\bibinfo {year} {2009})}\BibitemShut {NoStop}%
	\bibitem [{\citenamefont {Auld}(1990)}]{Auld1990v2}%
	\BibitemOpen
	\bibfield  {author} {\bibinfo {author} {\bibfnamefont {B.~A.}\ \bibnamefont
			{Auld}},\ }\href@noop {} {\emph {\bibinfo {title} {{Acoustic fields and waves
					in solids, Volume II}}}},\ \bibinfo {edition} {2nd}\ ed.\ (\bibinfo
	{publisher} {Robert E. Krieger Publishing Company},\ \bibinfo {address}
	{Malabar, Florida},\ \bibinfo {year} {1990})\BibitemShut {NoStop}%
	\bibitem [{COM()}]{COMSOL}%
	\BibitemOpen
	\href {www.comsol.com} {\emph {\bibinfo {title} {{COMSOL Multiphysics® v.
					5.4}}}},\ \bibinfo {organization} {{COMSOL AB}},\ \bibinfo {address}
	{Stockholm, Sweden}\BibitemShut {NoStop}%
	\bibitem [{\citenamefont {Yariv}(1973)}]{Yariv1973}%
	\BibitemOpen
	\bibfield  {author} {\bibinfo {author} {\bibfnamefont {A.}~\bibnamefont
			{Yariv}},\ }\bibfield  {title} {\enquote {\bibinfo {title} {Coupled-mode
				theory for guided-wave optics},}\ }\href@noop {} {\bibfield  {journal}
		{\bibinfo  {journal} {IEEE Journal of Quantum Electronics}\ }\textbf
		{\bibinfo {volume} {9}},\ \bibinfo {pages} {919--933} (\bibinfo {year}
		{1973})}\BibitemShut {NoStop}%
	\bibitem [{\citenamefont {Wolff}\ \emph {et~al.}(2015)\citenamefont {Wolff},
		\citenamefont {Steel}, \citenamefont {Eggleton},\ and\ \citenamefont
		{Poulton}}]{Wolff2015}%
	\BibitemOpen
	\bibfield  {author} {\bibinfo {author} {\bibfnamefont {C.}~\bibnamefont
			{Wolff}}, \bibinfo {author} {\bibfnamefont {M.~J.}\ \bibnamefont {Steel}},
		\bibinfo {author} {\bibfnamefont {B.~J.}\ \bibnamefont {Eggleton}}, \ and\
		\bibinfo {author} {\bibfnamefont {C.~G.}\ \bibnamefont {Poulton}},\
	}\bibfield  {title} {\enquote {\bibinfo {title} {Stimulated brillouin
				scattering in integrated photonic waveguides: Forces, scattering mechanisms,
				and coupled-mode analysis},}\ }\href@noop {} {\bibfield  {journal} {\bibinfo
			{journal} {Physical Review A}\ }\textbf {\bibinfo {volume} {92}},\ \bibinfo
		{pages} {013836} (\bibinfo {year} {2015})}\BibitemShut {NoStop}%
	\bibitem [{\citenamefont {Sohn}, \citenamefont {Kim},\ and\ \citenamefont
		{Bahl}(2018)}]{Sohn2018}%
	\BibitemOpen
	\bibfield  {author} {\bibinfo {author} {\bibfnamefont {D.~B.}\ \bibnamefont
			{Sohn}}, \bibinfo {author} {\bibfnamefont {S.}~\bibnamefont {Kim}}, \ and\
		\bibinfo {author} {\bibfnamefont {G.}~\bibnamefont {Bahl}},\ }\bibfield
	{title} {\enquote {\bibinfo {title} {Time-reversal symmetry breaking with
				acoustic pumping of nanophotonic circuits},}\ }\href@noop {} {\bibfield
		{journal} {\bibinfo  {journal} {Nature Photonics}\ }\textbf {\bibinfo
			{volume} {12}},\ \bibinfo {pages} {91} (\bibinfo {year} {2018})}\BibitemShut
	{NoStop}%
	\bibitem [{\citenamefont {Poveda}\ \emph {et~al.}(2019)\citenamefont {Poveda},
		\citenamefont {B{\"u}hler}, \citenamefont {S{\'a}ez}, \citenamefont
		{Santos},\ and\ \citenamefont {de~Lima~Jr}}]{Poveda2019}%
	\BibitemOpen
	\bibfield  {author} {\bibinfo {author} {\bibfnamefont {A.~C.}\ \bibnamefont
			{Poveda}}, \bibinfo {author} {\bibfnamefont {D.~D.}\ \bibnamefont
			{B{\"u}hler}}, \bibinfo {author} {\bibfnamefont {A.~C.}\ \bibnamefont
			{S{\'a}ez}}, \bibinfo {author} {\bibfnamefont {P.~V.}\ \bibnamefont
			{Santos}}, \ and\ \bibinfo {author} {\bibfnamefont {M.~M.}\ \bibnamefont
			{de~Lima~Jr}},\ }\bibfield  {title} {\enquote {\bibinfo {title}
			{Semiconductor optical waveguide devices modulated by surface acoustic
				waves},}\ }\href@noop {} {\bibfield  {journal} {\bibinfo  {journal} {Journal
				of Physics D: Applied Physics}\ }\textbf {\bibinfo {volume} {52}},\ \bibinfo
		{pages} {253001} (\bibinfo {year} {2019})}\BibitemShut {NoStop}%
	\bibitem [{\citenamefont {Waxler}\ and\ \citenamefont
		{Farabaugh}(1970)}]{Waxler1970}%
	\BibitemOpen
	\bibfield  {author} {\bibinfo {author} {\bibfnamefont {R.}~\bibnamefont
			{Waxler}}\ and\ \bibinfo {author} {\bibfnamefont {E.}~\bibnamefont
			{Farabaugh}},\ }\bibfield  {title} {\enquote {\bibinfo {title} {Photoelastic
				constants of ruby},}\ }\href@noop {} {\bibfield  {journal} {\bibinfo
			{journal} {J. Res. Nat. Bur. Stand.}\ }\textbf {\bibinfo {volume} {74}},\
		\bibinfo {pages} {215--220} (\bibinfo {year} {1970})}\BibitemShut {NoStop}%
	\bibitem [{\citenamefont {Shin}\ \emph {et~al.}(2013)\citenamefont {Shin},
		\citenamefont {Qiu}, \citenamefont {Jarecki}, \citenamefont {Cox},
		\citenamefont {Olsson}, \citenamefont {Starbuck}, \citenamefont {Wang},\ and\
		\citenamefont {Rakich}}]{Shin2013b}%
	\BibitemOpen
	\bibfield  {author} {\bibinfo {author} {\bibfnamefont {H.}~\bibnamefont
			{Shin}}, \bibinfo {author} {\bibfnamefont {W.}~\bibnamefont {Qiu}}, \bibinfo
		{author} {\bibfnamefont {R.}~\bibnamefont {Jarecki}}, \bibinfo {author}
		{\bibfnamefont {J.}~\bibnamefont {Cox}}, \bibinfo {author} {\bibfnamefont
			{R.}~\bibnamefont {Olsson}}, \bibinfo {author} {\bibfnamefont
			{A.}~\bibnamefont {Starbuck}}, \bibinfo {author} {\bibfnamefont
			{Z.}~\bibnamefont {Wang}}, \ and\ \bibinfo {author} {\bibfnamefont
			{P.}~\bibnamefont {Rakich}},\ }\bibfield  {title} {\enquote {\bibinfo {title}
			{{Tailorable stimulated Brillouin scattering in nanoscale silicon
					waveguides}},}\ }\href {\doibase 10.1038/ncomms2943} {\bibfield  {journal}
		{\bibinfo  {journal} {Nature Communications}\ }\textbf {\bibinfo {volume}
			{4}},\ \bibinfo {pages} {1944} (\bibinfo {year} {2013})}\BibitemShut
	{NoStop}%
	\bibitem [{\citenamefont {{Van Laer}}\ \emph {et~al.}(2015)\citenamefont {{Van
				Laer}}, \citenamefont {Kuyken}, \citenamefont {{Van Thourhout}},\ and\
		\citenamefont {Baets}}]{VanLaer2015}%
	\BibitemOpen
	\bibfield  {author} {\bibinfo {author} {\bibfnamefont {R.}~\bibnamefont {{Van
					Laer}}}, \bibinfo {author} {\bibfnamefont {B.}~\bibnamefont {Kuyken}},
		\bibinfo {author} {\bibfnamefont {D.}~\bibnamefont {{Van Thourhout}}}, \ and\
		\bibinfo {author} {\bibfnamefont {R.}~\bibnamefont {Baets}},\ }\bibfield
	{title} {\enquote {\bibinfo {title} {{Interaction between light and highly
					confined hypersound in a silicon photonic nanowire}},}\ }\href {\doibase
		10.1038/NPHOTON.2015.11} {\bibfield  {journal} {\bibinfo  {journal} {Nature
				Photonics}\ }\textbf {\bibinfo {volume} {9}},\ \bibinfo {pages} {199--203}
		(\bibinfo {year} {2015})},\ \Eprint {http://arxiv.org/abs/1407.4977}
	{arXiv:1407.4977} \BibitemShut {NoStop}%
	\bibitem [{\citenamefont {Wiederhecker}, \citenamefont {Dainese},\ and\
		\citenamefont {{Mayer Alegre}}(2019)}]{Wiederhecker2019}%
	\BibitemOpen
	\bibfield  {author} {\bibinfo {author} {\bibfnamefont {G.~S.}\ \bibnamefont
			{Wiederhecker}}, \bibinfo {author} {\bibfnamefont {P.}~\bibnamefont
			{Dainese}}, \ and\ \bibinfo {author} {\bibfnamefont {T.~P.}\ \bibnamefont
			{{Mayer Alegre}}},\ }\bibfield  {title} {\enquote {\bibinfo {title}
			{{Brillouin optomechanics in nanophotonic structures}},}\ }\href {\doibase
		10.1063/1.5088169} {\bibfield  {journal} {\bibinfo  {journal} {APL
				Photonics}\ }\textbf {\bibinfo {volume} {4}},\ \bibinfo {pages} {071101}
		(\bibinfo {year} {2019})}\BibitemShut {NoStop}%
	\bibitem [{\citenamefont {Eggleton}\ \emph {et~al.}(2019)\citenamefont
		{Eggleton}, \citenamefont {Poulton}, \citenamefont {Rakich}, \citenamefont
		{Steel},\ and\ \citenamefont {Bahl}}]{Eggleton2019}%
	\BibitemOpen
	\bibfield  {author} {\bibinfo {author} {\bibfnamefont {B.~J.}\ \bibnamefont
			{Eggleton}}, \bibinfo {author} {\bibfnamefont {C.~G.}\ \bibnamefont
			{Poulton}}, \bibinfo {author} {\bibfnamefont {P.~T.}\ \bibnamefont {Rakich}},
		\bibinfo {author} {\bibfnamefont {M.~J.}\ \bibnamefont {Steel}}, \ and\
		\bibinfo {author} {\bibfnamefont {G.}~\bibnamefont {Bahl}},\ }\bibfield
	{title} {\enquote {\bibinfo {title} {{Brillouin integrated photonics}},}\
	}\href {\doibase 10.1038/s41566-019-0498-z} {\bibfield  {journal} {\bibinfo
			{journal} {Nature Photonics}\ }\textbf {\bibinfo {volume} {13}},\ \bibinfo
		{pages} {664--677} (\bibinfo {year} {2019})}\BibitemShut {NoStop}%
	\bibitem [{\citenamefont {Safavi-Naeini}\ \emph {et~al.}(2019)\citenamefont
		{Safavi-Naeini}, \citenamefont {{Van Thourhout}}, \citenamefont {Baets},\
		and\ \citenamefont {{Van Laer}}}]{Safavi-Naeini2019}%
	\BibitemOpen
	\bibfield  {author} {\bibinfo {author} {\bibfnamefont {A.~H.}\ \bibnamefont
			{Safavi-Naeini}}, \bibinfo {author} {\bibfnamefont {D.}~\bibnamefont {{Van
					Thourhout}}}, \bibinfo {author} {\bibfnamefont {R.}~\bibnamefont {Baets}}, \
		and\ \bibinfo {author} {\bibfnamefont {R.}~\bibnamefont {{Van Laer}}},\
	}\bibfield  {title} {\enquote {\bibinfo {title} {{Controlling phonons and
					photons at the wavelength scale: integrated photonics meets integrated
					phononics}},}\ }\href {\doibase 10.1364/OPTICA.6.000213} {\bibfield
		{journal} {\bibinfo  {journal} {Optica}\ }\textbf {\bibinfo {volume} {6}},\
		\bibinfo {pages} {213} (\bibinfo {year} {2019})}\BibitemShut {NoStop}%
	\bibitem [{\citenamefont {{Van Laer}}, \citenamefont {Baets},\ and\
		\citenamefont {{Van Thourhout}}(2016)}]{VanLaer2016}%
	\BibitemOpen
	\bibfield  {author} {\bibinfo {author} {\bibfnamefont {R.}~\bibnamefont {{Van
					Laer}}}, \bibinfo {author} {\bibfnamefont {R.}~\bibnamefont {Baets}}, \ and\
		\bibinfo {author} {\bibfnamefont {D.}~\bibnamefont {{Van Thourhout}}},\
	}\bibfield  {title} {\enquote {\bibinfo {title} {{Unifying Brillouin
					scattering and cavity optomechanics}},}\ }\href
	{http://dx.doi.org/10.1103/PhysRevA.93.053828} {\bibfield  {journal}
		{\bibinfo  {journal} {Physical Review A}\ }\textbf {\bibinfo {volume} {93}},\
		\bibinfo {pages} {15} (\bibinfo {year} {2016})},\ \Eprint
	{http://arxiv.org/abs/1503.03044} {arXiv:1503.03044} \BibitemShut {NoStop}%
\end{thebibliography}

\appendix 

\section{Optomechanics of side-coupled waveguides}
\label{app:OM}

For intuition we adopt a quasi-static picture of the dynamics, treating the mechanical waves as stationary relative to light traveling along the waveguide.
At time \(t\), the mechanical wave deforms the waveguide uniformly along its length \(z\).
This deformation \(\U\prens{t,x,y} = \U_0\prens{x,y}\cos\Omega t\) of the waveguide's cross-section (radiation pressure) and the associated strain-induced change in the index (photoelastic effect)~\cite{Andrushchak2009} vary the permittivity, which to first order in \(\U\),
\begin{subequations}
\begin{align}
	\Delta\perm &= \delta_\U\perm \cdot \U\prens{t,x,y}& \\ \label{eq:deltaPerm}
													 &= \delta_{\partial\Omega}\prens{ \Delta\perm \proj_\perp- \perm\Delta\perm^{-1}\proj_\shortparallel\perm} \U\cdot\hat{n}	-\perm \frac{p \mathbf{S}}{\perm_0} \perm, &
\end{align}
\end{subequations}
shifts the wavevector of the guided optical wave \(\Delta\beta\cos\Omega t.\)  Here \(\delta_{\partial\Omega}\) is a delta function zero everywhere except dielectric boundaries, \(p\) is the photoelastic tensor, and \(S\) is the strain.  The projectors \(\proj_\perp\) and \(\proj_\shortparallel\) project out the field perpendicular and parallel to the normal \(\hat{n}\), respectively.  In this manner, the phase of light leaving the waveguide is acousto-optically modulated
\begin{equation} 
	a\prens{t,z = L} = a\prens{0,L}e^{i\Delta\beta L\cos\Omega t}.
	\label{eq:phaseMod}
\end{equation}

The phase modulation index \(h_\trm{oa} \equiv \Delta\beta L\) is related to the optomechanical coupling \(\gsquig\) familiar in stimulated Brillouin scattering.
By adopting a power-orthogonal basis for the electric field \(\E = \Sigma_n \E_n a_n e^{-i\prens{\omega + n\Omega}t}\), the interaction can be expressed in terms of coupled-modes~\cite{Yariv1973,Wolff2015}.
For short lengths and low RF drive powers, the \(n = +1\) mode evolves as
\begin{equation}
	\prens{\partial_z + v_\trm{g}^{-1}\partial_t} a_1\prens{t,z} = i\gsquig b a_{0}
	\label{eq:coupledModes}
\end{equation}
where \(b\) is the amplitude of the mechanics \(\U = \U_0 b e^{-i\Omega t}\).
The optomechanical coupling can be expressed in terms of the mode profiles
\begin{equation}
	\gsquig = -\frac{\omega}{\mathcal{P}_{1}}\int\trm{d}A \E^*_1\delta_u\perm\cdot \U_0 \E_0 
	\label{eq:gsquig}
\end{equation}
where \(\mathcal{P}_i\) is the time-averaged optical power into the waveguide for mode \(i\). If we use a power-\emph{orthonormal} basis for the optics such that \(\left|a_i\right|^2\) is the power in mode \(i\), \(\mathcal{P}_i\) becomes \(1\) for all \(i\) and Eq.~\ref{eq:gsquig} takes on a more symmetric form for various mode pairs. 

We have yet to choose a normalization for the displacement \(\U_0\) and thereby units for both \(\gsquig\) and \(b\). For the devices described here, the devices in Sohn~\emph{et al.}~\cite{Sohn2018}, and work on AO-modulated Mach-Zehnder interferometers~\cite{Poveda2019,Cai2019,Khan2019}, the mechanical wave propagates across the waveguide. In this configuration, the displacement of the waveguide scales as the square root of the mechanical power density, which is the power per unit length along the waveguide. A \SI{1}{\milli\watt} wave generated by a \SI{100}{\micro\meter}-wide IDT will deform the waveguide the same as that of a \SI{2}{\milli\watt} wave from a \SI{200}{\micro\meter}-wide IDT. We normalize the displacement field \(\U_0\) such that \(b^2\) is the mechanical power density with units of \(\si{\watt}/\si{\meter}\). Therefore the coupling coefficient \(\gsquig\) has units of \(1/\sqrt{\si{\watt}\cdot\si{meter}}\). 

Finally, when \(\gsquig b \ll L\), we can integrate Equation~\ref{eq:coupledModes} and compare it to equation Equation~\ref{eq:phaseMod} to find
\begin{subequations}
\begin{align}
	\frac{h_\trm{oa}}{2} &= \gsquig b L  \\
	&= \gsquig \sqrt{P_\trm{m} L}.
\end{align}
\end{subequations}
Substituting in the efficiency of the IDT, \(P_\trm{m} = \left|\tbmu\right|^2 P\) where \(P\) is the RF power incident on the IDT, we arrive at Equation~\ref{eq:modulationIndex}.

It may be surprising that the modulator's efficiency \(h_\trm{oa}^2/4 P\) --- the sideband power ratio divided by the RF drive power, which is the square of the expression in Equation~\ref{eq:modulationIndex} --- scales as \(L\) and not \(L^2\), but this scaling is essentially the same as in electro-optic modulation. A typical electro-optic modulator is nearly identical to the optomechanical modulator described here except that instead of applying a displacement \(\U\) in Equation~\ref{eq:deltaPerm} to change the permittivity of a waveguide, a voltage is applied to a capacitor to shift the effective index \(\Delta n_\trm{eff}\) by the electro-optic effect. The phase shift of light transmitted through the waveguide \(\Delta n_\trm{eff} L\) is proportional to \(L\), and therefore the power in the sidebands is proportional to \(L^2\). But just as the total mechanical power \(P_\trm{m}\) scales as \(L\), the energy in a capacitor along the waveguide scales as \(L\). The result is that, like the optomechanical modulator described here, the efficiency of an electro-optic modulator scales as \(L\).

Equation~\ref{eq:gsquig} is used to compute the \(\gsquig\) plotted in Figure~\ref{fig:AO} of the main text. In our calculation we model the photoelastic effect in sapphire using coefficients measured in ruby~\cite{Waxler1970} and, for the radiation pressure term, LN~is treated as an isotropic medium with refractive index of \(2.15\).

Finally, our device is at heart an electrically-driven analog of the optomechanical waveguides investigated in the field of guided-wave Brillouin scattering. In those systems \cite{Shin2013b,VanLaer2015,Wiederhecker2019,Eggleton2019,Safavi-Naeini2019} the mechanical motion is typically driven optically, while here the mechanical motion is driven electrically. The optomechanical coupling coefficient $\gsquig$ used here (Equation~\ref{eq:gsquig}) is, up to a few conversion factors, identical in nature to the Brillouin gain coefficient $\mathcal{G}_{\text{B}}$ and the refractive-index sensitivity to mechanical motion $\partial_{x}n_{\text{eff}}$. The explicit connection is as follows. The modulation index can be written as
\begin{equation}
    \frac{h_\trm{oa}}{2} = \gsquig b L = \frac{\Delta \beta L}{2} = k_{0} \partial_{x}n_{\text{eff}} L x
\end{equation}
with $k_{0}$ the free-space optical wavevector and $x$ a coordinate representing the mechanical motion. From previous work \cite{VanLaer2015,VanLaer2016} we have 
\begin{equation}
    \partial_{x}n_{\text{eff}} = c \sqrt{\frac{\mathcal{G}_{\text{B}} \omega_{\text{m}}^{2} m_{\text{eff}}}{2\omega_{0}Q_{\text{m}}}}
\end{equation}
with $c$ the speed of light, $\omega_{0}$ the optical frequency, $Q_{\text{m}}$ the mechanical quality factor, $\omega_{\text{m}}$ the mechanical frequency, $m_{\text{eff}} = \int\trm{d}A \rho |\U|^{2}$ the effective mechanical mode mass, and $\rho$ the mass density. This shows that the optomechanical coupling coefficient $g$ used here and the Brillouin gain coefficient are rigorously connected. We note that the optomechanical coupling coefficient $g \propto \sqrt{\mathcal{G}_{\text{B}}}$ scales as the square root of the Brillouin gain coefficient, since the latter takes into account both optical driving and optical read-out of the mechanics, whereas in this work only the read-out of the mechanics occurs optically.

\section{Regressing the photoelastic coefficients}
\label{app:regression}

Our measurements of \(\gsquig\) in Figure~\ref{fig:AO} show rough agreement with the coupling predicted using values of the photoelastic tensor of bulk LN measured by Andrushchak~\emph{et al.}~\cite{Andrushchak2009}. We would like to go beyond this qualitative comparison and find the components \(p_{ij}\) that best fit the data. There are \(d = 8\) distinct components of \(p\) for LN, with the remaining components constrained by the symmetry of the crystal lattice. In order to avoid overfitting our $n = 14$ measurements --- 2 modes, Rayleigh and SH, measured at 7 angles \(\theta_\trm{x}\), each --- we start with a regularized fit before reducing the regression to just three components of \(p\).

In Appendix~\ref{app:OM}, we outline how the photoelastic tensor \(p\) is related to the coupling coefficient \(\gsquig\). Since \(\gsquig\) is proportional to \(\delta\varepsilon_u\) and \(\delta\varepsilon_u\) is linearly dependent on \(p\), the coupling is a linear function of \(p\) which can be expressed
\begin{equation}
    \gsquig = \left| A p + b \right|
    \label{eq:gsquigFit}
\end{equation}
In Equation~\ref{eq:gsquigFit}, \(\gsquig \in \mathbb{R}^n\), \(p\in\mathbb{R}^d\), \(A \in \mathbb{C}^{n\times d}\), and \(b\in\mathbb{C}^n\).  Both \(A\) and \(b\) are computed as described in the main text. They are complex because \(u\), the mechanical field, is complex. The phase of \(u\) can be chosen to make \(\gsquig\) real but this in turn affects the phase of \(\tbmu\). Although the phase of \(\gsquig\tbmu\) does factor into the phase of \(S_\trm{ao}\), we restrict our measurements to the magnitude of \(\gsquig\tbmu\).

We start by solving
\begin{align}
 \min_p   \left\| \left| Ap + b \right| -  \gsquig \right\| + \lambda \left\| p - p_0\right\|_{1}
	\label{eq:regularizedFit}
\end{align}
where \(\left\| \cdot \right\|_1\) is the L$^1$ norm. The second term in Equation~\ref{eq:regularizedFit} is used to regularize the fit. It penalizes deviation from \(p_0\), the values measured for bulk LN by Andrushchak~\emph{et al.}. Using an L$^1$ norm encourages \(p - p_0\) to be sparse. As \(\lambda\) is increased, some components of \(p - p_0\) fix to \(0\). At first, the quality of the fit (the first term in the right-hand side of Equation~\ref{eq:regularizedFit}) is relatively unaffected by the reduction of dimension \(d\).  Ultimately the regularization term begins to exert a pressure on the components of \(p\) needed to fit the data and the fit diverges from the measured values. At this point we are left with just three components of \(p\) which deviate from \(p_0\) : \(p_{33}\), \(p_{44}\), and \(p_{41}\) in Voigt notation. 

\begin{table}[h]
	\begin{tabular}{| r | c | c |}
		\hline
		 & Andrushchak~\emph{et al.}~\cite{Andrushchak2009} & Fit \\
		\hline
		$p_{11}$ & $-0.021 \pm 0.018$  &  ---  \\ 
		\hline
		$p_{12}$ & $0.060 \pm 0.019 $  &  ---  \\ 
		\hline
		$p_{13}$ & $0.172 \pm 0.029 $  &  ---  \\ 
		\hline
		$p_{31}$ & $0.141 \pm 0.017 $  &  ---  \\ 
		\hline
		$p_{33}$ & $0.118 \pm 0.020 $  &  0.038  \\ 
		\hline
		$p_{14}$ & $-0.052 \pm 0.007$ &  ---  \\ 
		\hline
		$p_{41}$ & $-0.109 \pm 0.017$ & -0.038  \\
		\hline
		$p_{44}$ & $0.121 \pm 0.019$ &  0.085  \\
		\hline
	\end{tabular}	
	\caption{Photoelastic coefficients for bulk LN and the best fit from the \(\gsquig\) measurements in Figure~\ref{fig:AO}.}
	\label{tab:photoelastic}
\end{table}

We repeat the regression setting \(\lambda\) to 0 and removing all other components of \(p\) from the fit to find the values in Table~\ref{tab:photoelastic} giving us the dashed curves in Figure~\ref{fig:AO}b.

\end{document}